\documentclass[12pt]{article}
\usepackage{graphicx} 
 \usepackage{color}
 \usepackage{mdframed}
 \usepackage{amsmath}
 \usepackage{multirow}
\usepackage[final]{pdfpages}
 
\title{Impact of the sodium and calcium chlorides uptake on the interfacial behavior of ice: premelting, structure, and dynamics.}
\author{{\L}ukasz Baran$^{1,2}$, Luis G. MacDowell$^{2}$}
\date{$^1$ Department of Theoretical Chemistry, Institute of Chemical Sciences, Faculty of Chemistry, Maria Curie-Sklodowska University in Lublin, Poland \\
$^2$ Departamento de Qu\'{\i}mica F\'{\i}sica, Facultad de Ciencias Qu\'{\i}micas, Universidad Complutense de Madrid, Spain}
\usepackage[top=2.5cm, bottom=1.5cm, left=2.5cm, right=2.5cm]{geometry}

\begin{document}
\maketitle

\clearpage

\begin{abstract}
    {\bf Hypothesis:} Seawater ice and frozen aqueous solutions in contact with air can exhibit a thin quasi-brine surface layer intruding between ice and vapor, but a detailed characterization of surface properties and its relation to three phase coexistence has been lacking. Using thermodynamic arguments we show how it is possible to  characterize the surface layers by comparison to the three phase ice-brine-air bulk phase diagram, despite the difficulty to control or monitor all of the relevant thermodynamic fields of the two component system. \\
    
\noindent
 {\bf Simulations:} We performed computer simulations of surface briny layers of sodium and calcium chloride adsorbed on ice. Using suitable order parameters and a rigorous geometrical dividing surface,  we are able to characterize the layer's thermodynamic state, measure its properties and relate them to the corresponding  properties of the bulk solution. \\

\noindent 
 {\bf Results:}
Our results confirm that undersaturated briny surface layers can form down to the eutectic point, with a maximum concentration that is bound by the liquidus line of the ice-brine phase diagram. 
Such layers are distinct from finite size realizations of three phase coexistence,
and can be regarded as genuine surface states, but their salt content can increase the premelting layer thickness by a factor of two or more. Owing to this significant thickness, these layers can be related to bulk electrolyte solutions of similar concentration, both as regards the structural organization of ions and the dynamical  properties of the quasi-liquid film. 
\end{abstract}

\section{Introduction}

Hexagonal ice is the most commonly observed form of solid water on  Earth, covering over 10\% of its surface \cite{bartels12}. On the earth crust it dictates major geological processes, such as glacier sliding or frost heaving \cite{dash06}. On the atmosphere, it plays a fundamental role in  precipitation, atmospheric chemistry, the radiation budget or climate change \cite{magnan21}. Importantly, ice also acts as a reactive medium that mediates the transport of contaminants between the atmosphere and the cryosphere \cite{bartels14}. Consequently, the understanding of the interfacial properties  where this exchange takes place is of outmost importance. Yet, the microscopic  structure of the ice interface and its dependence on thermodynamic conditions have been the matter of long standing debates \cite{dash06, dash95} and pose  challenges on our understanding of fundamental concepts in  chemical-physics \cite{petrenko99}.

It is now widely accepted that, below the melting point, a disordered layer of premelted ice appears at the ice-vapor interface. This layer, commonly referred to as the quasi-liquid layer (QLL), exhibits properties similar to those of supercooled liquid water under the same thermodynamic conditions \cite{kling18, louden18, nagata19}. However, there is no consensus regarding either the temperature at which premelting begins or the extent to which it occurs. Reported onset temperatures range from as low as $T=200$~K \cite{wei01}, to the melting point itself \cite{furukawa87} or anywhere in between \cite{slater19}, with layer thicknesses varying from a few Angstroms up to hundreds of nanometers \cite{bartels14, slater19}. Such discrepancies likely arise from differences in the surface sensitivity of experimental techniques \cite{bartels14, slater19}, the presence of impurities \cite{mcneill06, mcneill12, elbaum93}, and whether the measurements were conducted at \cite{qiu18, sanchez17} or away from ice-vapor coexistence \cite{murata16, asakawa16}. Nevertheless, both computer simulations \cite{llombart20, baran24b, conde08} and recent experimental studies are in agreement that at 250~K, a significant premelting layer has been formed \cite{sanchez17} and that its thickness can reach the nanometer scale close to the triple point \cite{bluhm02,mitsui19}.  


Recently the long-standing view of the QLL as a thin layer of meltwater has been challenged. Demmenie \textit{et al.} \cite{demmenie25} proposed that the observed liquid-like layer may instead consist of deposited water vapor rather than a liquid film. Their conclusions stem from measurements of non-zero contact angles at the melting point, consistent with theoretical predictions based on Hamaker constants \cite{baran24b, luengo22}. The absence of complete wetting at the triple point implies that the layer cannot be considered a liquid in the strict thermodynamic sense but rather a two-dimensional mobile gas. Regardless of the correct interpretation, it is evident that despite the extensive attention the ice interface has received since Faraday’s famous regelation experiments \cite{faraday60}, its complete and unambiguous characterization remains elusive.

Since the pure-water conditions are still a matter of  debate, it is not surprising that the clear picture of how the impurities affect the interfacial behavior of ice is also yet to be settled. Chemical compounds, such as strong acids (H$_2$SO$_4$ or HNO$_3$), are present in polar stratospheric and subtropical cirrus clouds that are formed at temperatures as low as $190$~K \cite{mohler05, popp04} and are crucial for the understanding of the depletion of ozone layer \cite{heckendorn09, solomon11}. At higher temperatures, organics \cite{mcneill12, singh01}, different salts \cite{sivells23}, and other contaminants are present in the ice samples, while briny water droplets are easily formed in the atmosphere from sea sprays \cite{bartels12}. Establishing their influence on ice behavior both in urban and natural environments is therefore crucial for understanding the atmospheric processes and climate change. 

Adsorption of ions and other impurities can lead to the formation of brine on the topmost layers of environmental snow and ice. In bulk, the addition of a second component causes a freezing point depression, driving the system closer to melting conditions. As a result, one expects that surface impurities could induce further premelting in order for the quasi-liquid layer  to reach the equilibrium bulk concentration of brine in coexistence with ice.  This behavior has been reported for ions \cite{mcneill06, berrens22, hundait17}. In particular, Cho \textit{et al.} reported that the concentrations of aqueous NaCl layers formed on the ice surface  are well described by the bulk equilibrium brine concentration along the liquidus NaCl-H$_2$O phase line up to the eutectic point \cite{cho02}, a trend also observed for seawater ice \cite{richardson76}. On the other hand, several experimental measurements show that hydrophilic volatile organic compounds such as acetic acid and acetone do not increase the thickness of interfacial layers \cite{starr11, krepelova13}. However, the measurements were performed at extremely low temperatures where the extent of premelting is already very small, raising doubts whether similar results would hold at higher temperatures. Interestingly, brines have also been found to exist below the eutectic temperature which may arise from the metastability of such phases \cite{gough11, toner14} or from the curvature-induced melting at the grain boundaries described by the Gibbs-Thomson equation \cite{aristov97, christenson01}. 

Freshwater and seawater contain different amounts of ions. Therefore, the premelting behavior at the air-freshwater ice surface is significantly different from that at air-seawater ice interfaces. Kahan \textit{et al.} reported that the QLL formed in the latter case can be meaningfully described in terms of bulk seawater at the same thermodynamic conditions \cite{kahan14}. Since sea ice is the most abundant form of ice found on Earth, it is the most relevant one to be studied from the environmental perspective. Although regular seawater has a salinity not exceeding 35 g of salt per 1 kg of water, freezing expels brine from the ice, forming concentrated droplets that can be trapped at the interface of the grain boundaries \cite{ingolf75, shcherbina03, peterson18, crabeck19}. These sea ice brine pockets are an environment for microbes that can withstand extreme hypersaline conditions \cite{junge01}, making them relevant not only for understanding Earth’s cryosphere but also as analogs for potential extraterrestrial life on icy moons \cite{buffo20, vance21}. Of course, the characterization of premelting layers of seawater ice is in principle very difficult, because seawater is a complex mixture of several many salts. In practice, however, sodium chloride accounts for 85\%  of its mole fraction, and several of the remaining salts freeze out at significantly higher temperatures than sodium chloride. 

In special conditions, salty ice-air interfaces can be formed from hypersaline brines completely different from seawater. Such is the case of the Don Juan Pond, in the Antarctica, which is mostly formed from calcium chloride \cite{dickson13}, or the Gaet’ale Pond  in Ethiopia,  which owing to the additional presence of magnesiunm chloride holds the world's hypersaline record at 433 g/kg concentration. \cite{perez17}

In this study we perform computer simulations to investigate how the adsorption of sodium and chloride ions affects the premelting behavior of an ice sample, as a proxy for seawater ice interfaces.  To explore properties of premelting films in extreme conditions, as  e.g. in Don Juan Pond,  we also examine premelting layers with adsorbed calcium chloride, a divalent salt.

Our results show that the extent of premelting varies significantly with temperature and surface coverage, but its thickness is always larger than that on purewater ice. The concentration of the resulting films is roughly bound by the equilibrium brine concentration along the liquidus line, a maximum that is reached as the films grow thicker. However,  thin premelting films can be equilibrated which exhibit significantly smaller salinity than the equilbrium brine concentration in bulk.
A detailed analysis reveals that the water molecules in these layers exhibit suppressed motion as compared to the pure-water bulk system, but the rheological properties are similar to those found in bulk aqueous electrolyte solutions of similar concentration. Our study hinges on the thermodynamic distinction between surface premelting and bulk ice-brine coexistence, and shows how the properties of the confined brine are related to those of the bulk solution.

\section{Thermodynamics of briny ice surface melting}\label{sec:thermo}

\begin{figure}
    \centering
    \includegraphics[width=0.75\linewidth]{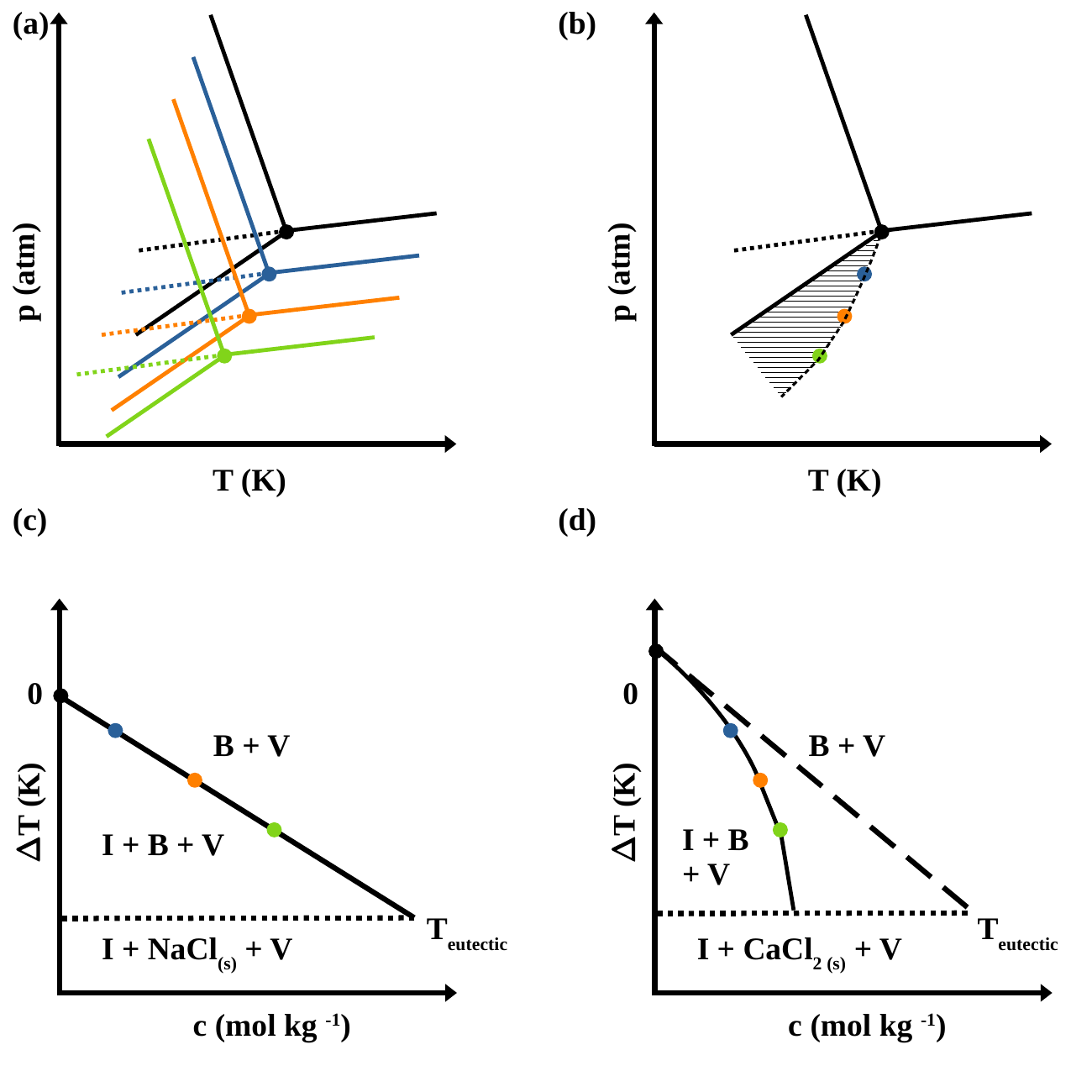}
    \caption{Schematic phase diagram of salt-water solutions in the proximity of the triple point. Part (a): Full black lines denote the phase boundaries of pure water, showing the melting, condensation and sublimation boundaries. The black dotted line denotes the metastable prolongation of the liquid-vapor condensation line. Similar pseudo one-component phase diagrams can be drawn for the phase coexistence of brines of fixed composition, denoted here in blue, orange and green in order of increasing concentration. Filled circles denote the corresponding ice-brine-vapor triple points. Joining this points together produces a line of ice-brine-vapor triple points. Part (b): Binary phase diagram for the ice-salt phase coexistence. The shaded area between the pure-component sublimation line and the triple-point line indicates the loci of thermodynamic states  where the equilibrium ice-vapor coexistence is possible. 
    Parts (c) and (d): Examples of equilibrium phase diagram for a two-component system consisting of water and a soluble salt - "ideal" (as in NaCl) (c) and "non-ideal" (as in CaCl$_2$) salts (d). The marked points on the liquidus line correspond to the triple points shown in parts (a) and (b). The abbreviations are: ice (I), brine (B), and vapor (V).}
    \label{fig:diagram}
\end{figure}

In a one component system, equilibrium premelting refers to the situation where bulk ice is exactly at coexistence with its bulk vapor at a fixed temperature on the sublimation line (Figure~\ref{fig:diagram}-a). 
At such conditions, the outermost water molecules on the solid surface of the ice-vapor interface are under-coordinated and can melt at temperatures below the bulk triple point, leading to the formation of a thin liquid-like layer. The thickness of the liquid film so adsorbed may be characterized in terms of the distance of the sublimation line away from the  metastable prolongation of the liquid-vapor coexistence line, shown with dashed lines in Figure~\ref{fig:diagram}-a. Now, the task is to somehow measure the thickness of this layer $h(T)$ as the equilibrium ice-vapor surface is heated along the sublimation line. Two possible outcomes are expected as the triple point is approached. Either $h(T)$ gradually diverges and becomes infinitely large, corresponding to  a case of  \textit{complete surface melting} (water wets ice); or it converges to some finite thickness, corresponding to a situation of  \textit{incomplete surface melting} (water does not completely wet ice). The current understanding of the problem, as evidenced from a combination of experiments, computer simulations and theoretical methods is that pure ice exhibits incomplete surface melting \cite{macdowell25}.

The characterization of equilibrium premelting films becomes far more complex with the addition of the second component, as the thermodynamic space now takes an extra degree of freedom. In practice, this means that 
two-phase coexistence is no longer characterized by the temperature alone, but requires an additional thermodynamic field which is difficult to control or monitor in practice. 

Theoretically, the characterization of premelting for pure substances could be extended to mixtures by using a reference solution of fixed salt concentration in place of the pure water reference liquid. For salts that are soluble in liquid, but are hardly soluble in the vapor and solid phases (such as NaCl and CaCl$_2$), the free energy of the liquid solution decreases by mixing, while that of ice and vapor remains hardly unchanged. The resulting melting point depression leads to a shift of the melting line to lower temperatures, while the corresponding boiling point elevation shifts the condensation line to lower pressures. Intersection of these two lines then leads to a new triple point for the reference brine solution. Starting from this triple point, it is possible to 
draw a new sublimation line for ice and vapor phases coexisting at equal chemical potential and temperature as the metastable reference brine solution of fixed concentration.  The result of this construction is a new set of two phase lines, equal to those of the pure substance, but shifted to lower temperature and pressure (depicted as colored lines in  Figure~\ref{fig:diagram}-a). 

Repeating this construction for all possible brine concentrations draws a surface on the $p,T$ phase diagram corresponding to the loci of points where equilibrium ice-vapor coexistence is possible. This region is bounded from above by the sublimation line of pure water, and bounded from below by a line of ice-brine-vapor triple points (shaded area in Figure~\ref{fig:diagram}-b) that terminates at the eutectic point. An equilibrium premelting film can exist for any thermodynamic state within this region. 

In principle, therefore, an analysis of the premelting film thickness could be done in analogy with that for pure substances, by monitoring the film thickness as a function of temperature, $h(T,C_0)$ for a solid-vapor phase line with equal chemical potential as a reference brine solution of known concentration, $C_0$. Unfortunately, in practice this is not possible, as it corresponds to moving along a trajectory of unknown chemical potentials. The alternative, i.e., monitoring $h(T,p)$ as a function of the vapor pressure is not convenient either.
Ice and vapor in equilibrium with a solution have an exceedingly small concentration of salt, so that the  ice-brine-vapor triple line runs below but exceedingly close to the sublimation line of pure ice-vapor equilibrium. Therefore, the full range of equilibrium premelting films at constant temperature are actually realized in an extremely small range of pressures that would be difficult to measure.

It follows that monitoring the equilibrium structure at the ice-vapor interphase of a two component system is very difficult to do in terms of the two bulk thermodynamic variables required to fully characterize the actual two phase system. 

This turns to be extremely inconvenient. Without control of the extra degree of freedom, it is not possible a priori to tell whether the ice-brine-vapor system is an actual realization of interfacial premelting at ice-vapor coexistence, or a small system size realization of three phase bulk coexistence.

The only feasible alternative is  to prepare the two component solid-vapor interface of unknown chemical potential with some fixed known amount of surface ions, and let the interface relax to an equilibrium surface state of fixed salt adsorption, $\sigma$. The premelting film thickness, $h(T,\sigma)$ then becomes a function of temperature and $\sigma$ alone.

The expectations from this method of preparation is that at a given fixed temperature, premelting films can be prepared ranging from zero salt concentration, to 
the equilibrium brine concentration at that temperature's triple point.  On the phase diagram of Figure~\ref{fig:diagram}-b, the first case corresponds to a state on top of the sublimation line of pure water (i.e.,
the black line bounding the shaded area from above); the second case corresponds to a point on the ice-brine-vapor triple point (i.e. the dashed line bounding the shaded area from below). In this limit, two outcomes are possible. Either the premelting layer remains finite, at some unknown surface concentration; or  the ice surface becomes completely wet by the brine (i.e. brine-ice surface melting), and there is no longer a distinction between the (thick) premelted layer and the three phase bulk coexistence.

This discussion shows that detailed knowledge  of the ice-brine-vapor phase diagram is a prerequisite for a proper characterization of the interfacial physics. Fortunately, in this case the low solubility of salts in the solid and vapor phases is very helpful. The ice-brine equilibrium at constant atmospheric pressure is well documented, and for "ideal" salts, such as NaCl, looks qualitatively similar to that depicted in Figure~\ref{fig:diagram}-c, whereas for "non-ideal" salts, such as CaCl$_2$, it looks as in Figure~\ref{fig:diagram}-d. In either case,  the diagram illustrates the phase coexistence of a brine solution of given temperature in equilibrium with almost pure ice. By moving the two phase system to lower pressure, ice will eventually reach a state where it also achieves phase equilibrium with almost pure vapor. The pressure drop required to achieve this is so small, that the properties of the brine in equilibrium with ice will have hardly changed.  Whence, the constant pressure two phase coexistence phase diagram is virtually the same as that corresponding to three phase coexistence, and the liquidus line may also be interpreted as a very good proxy for a three phase line indicating the composition of brine in simultaneous equilibrium with ice and vapor.

With the help of the bulk phase diagram, it now becomes possible to characterize the surface equilibrium states and distinguish genuine surface states from finite size realizations of the triple point. In our work, we will show how to relate the surface coverage to an effective film concentration, which serves as a reference to discriminate between bulk and surface effects. Indeed, the bulk phase diagram constrains equilibrium triple points to have a brine concentration equal to the liquidus concentration. Accordingly, films with nominal concentration smaller than this maximum, can be safely identified as surface states, and distinguished from bulk three phase coexistence.

In summary, this lengthy and intricate discussion shows that the study of premelting films in two components systems is considerably more difficult than for a one component system. Particularly, the surface states cannot be properly characterized without a detail knowledge of the bulk phase diagram. In the next sections, we will use this essential information in order to  characterize the surface properties  of quasi-liquid briny ice films under a suitable thermodynamic framework.

\section{Computational methods}
\subsection{Model and sample preparation}
Water is modeled using  TIP4P/Ice \cite{abascal05}, while the ions are described with the Madrid-2019 force field \cite{zeron21}. Although the latter was initially parametrized for the TIP4P/2005 water model, its compatibility with the Ice version has been recently demonstrated \cite{blazquez24}. The interaction parameters are shown in Table~\ref{tab:model}. For both models, dispersion interactions are truncated at 1 nm. 

Notice that the Madrid-2019 force field is a scaled charge model, where the formal ionic charges are scaled by a factor of $0.85$. This provides excellent agreement with experimental results for the structural properties but leads to significant deviations in dynamical behavior. The latter properties can be improved by further scaling, but this is at the expense of deteriorating structural quantities \cite{blazquez23}. Moreover, we have recently shown that charge scaling may not be necessary at all for reproducing structural properties, provided the water model reproduces a correct value of the dielectric constant \cite{baran25}. Nevertheless, for the present study of ice premelting, we employ the water model exhibiting a melting point which is close to the experimental value, while keeping in mind the possible effects of charge scaling on the results.

\begin{table}[h!]
    \centering
    \begin{tabular}{ccccc}
        LJ Interaction & $\sigma$~(\AA) & $\varepsilon$~(kcal/mol) & Charge &  q (e)  \\
        \hline
        \hline
         Na$^+$-Na$^+$  &  2.21737 & 0.351902 & Na$^+$ & $+0.85$ \\
         Ca$^{2+}$-Ca$^{2+}$ & 2.66560 & 0.121224 & Ca$^{2+}$ & $+1.70$ \\
         Cl$^-$-Cl$^-$  &  4.66906 & 0.018385 & Cl$^-$ & $-0.85$\\
        \hline 
         Na$^+$-Cl$^-$  &  3.00512 & 0.343904 &  \\
         Ca$^{2+}$-Cl$^-$  & 3.15000  & 0.239006 &  \\
        \hline
         Na$^+$-O       &  2.60838 & 0.189624 & \\
         Ca$^{2+}$-O$^-$  & 2.40000  & 1.732792 &  \\
         Cl$^-$-O       &  4.23867 & 0.014814 & \\
         O-O            &  3.16680 & 0.210840 &  O   & 0.0 \\
                        &         &         &  H   & $+0.5897$ \\ 
                        &         &         &  M   & $-1.1794$ \\
        \hline
        \hline
    \end{tabular}
    \caption{Madrid-2019 NaCl and CaCl$_2$ parameters combined with TIP4P/Ice water model taken from Ref.~\cite{blazquez24}.}
    \label{tab:model}
\end{table}

\begin{figure}
    \centering
    \includegraphics[width=0.75\linewidth]{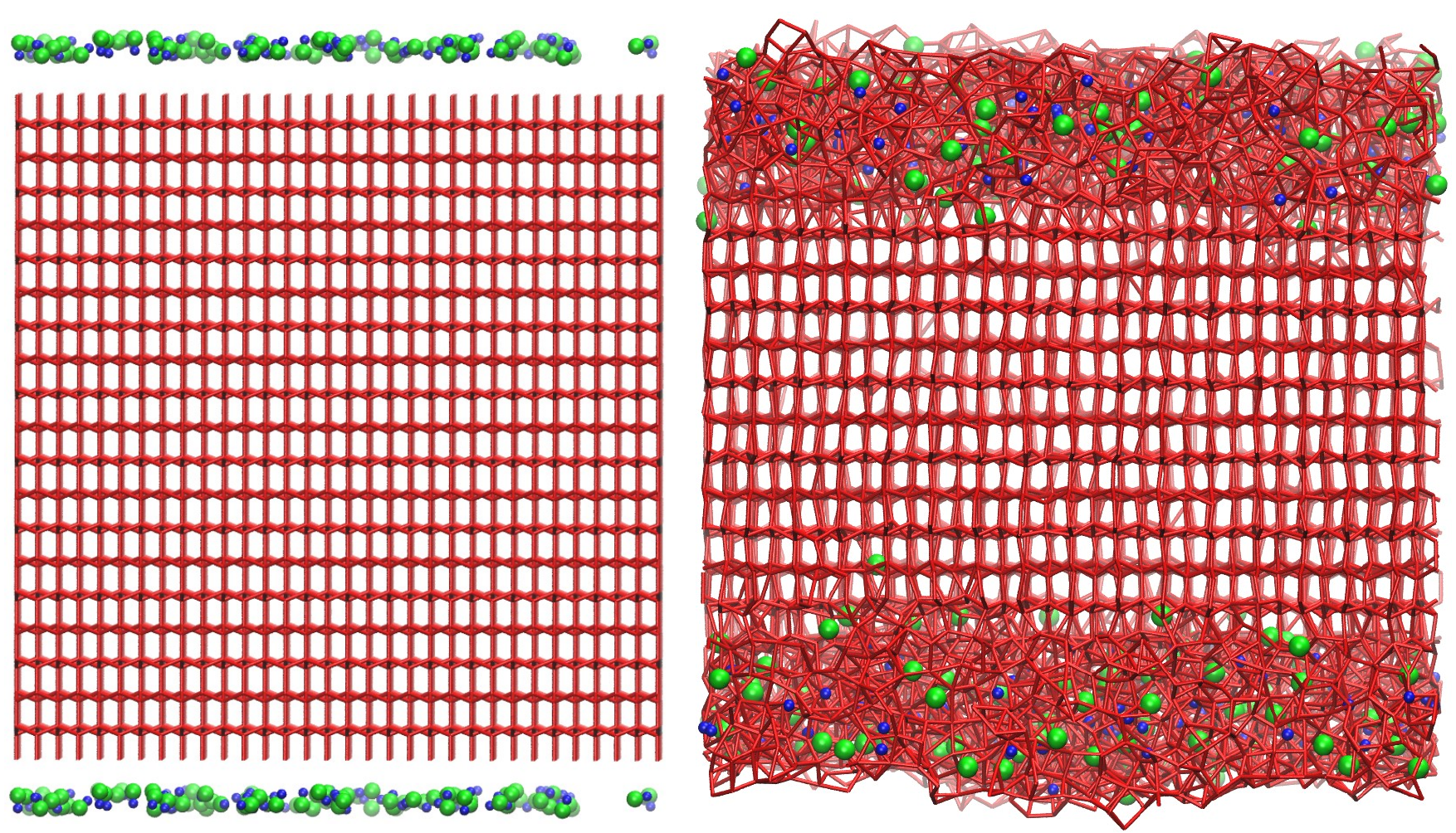}
    \caption{Snapshot of the simulation setup for $\sigma=2.2$ at $T=260$~K. Left panel: an initial configuration of a perfect ice crystal (red sticks) with the basal plane exposed to the vapor with the Na$^+$ (blue spheres) and Cl$^-$ (green spheres) ions deposited slightly above the interface. Right panel: equilibrated configuration after 400 ns of simulation time.}
    \label{fig:init}
\end{figure}

\begin{table}
    \centering
    \scalebox{0.8}{
    \begin{tabular}{cccccccc}
         System & N$_{\text{water}}$& N$_{\text{NaCl}}$ & N$_{\text{Ca$^{2+}$}}$ & N$_{\text{Cl$^{-}$}}$ & $\sigma$ (salt/nm$^2$) & $L_x\times L_y\times L_z$ ~(\AA$^3$)\\
         \hline
         \hline
        \multirow{ 4}{*}{ice-vapor} & \multirow{ 4}{*}{10240} & - & - & - & - &\multirow{ 4}{*}{$62\times72\times170$}\\
         & &  25 & 25 & 50 & 0.55 &  \\
         & &  50 & 50 & 100 & 1.1 &  \\
         & &  98 & 100 & 200 & 2.2 &  \\
        \hline
        \hline
        System & N$_{\text{water}}$&N$_{\text{NaCl}}$ & N$_{\text{Ca$^{2+}$}}$ & N$_{\text{Cl$^{-}$}}$ & $c$ (mol/kg) & $L_x\times L_y\times L_z$ ~(\AA$^3$) \\ 
        \hline
        \hline
         bulk ice   &  1280 & - & - & - & - & $36\times32\times36$ \\
         \hline
         bulk water & \multirow{ 6}{*}{2220} & - & - & - & - & \multirow{ 6}{*}{$40\times40\times40$} \\
        \multirow{5}{*}{bulk solution}& & 20 & 20 & 40 & 0.5 \\ 
        & & 40 & 40 & 80  & 1.  & \\
        & & 80 & 80 & 160 & 2. & \\
        & & 120 & - & - & 3. & \\
        & & 160 & - & - & 4. & \\
        \hline
        \hline
    \end{tabular}
    }
    \caption{Number of water molecules and the system size used for the ice-vapor (bulk) simulations with a different number of Na$^+$, Ca$^{2+}$, and Cl$^-$ ions and corresponding surface coverage, $\sigma$ (molality, $c$). The reference system sizes are also given. }
    \label{tab:gamma}
\end{table}

For each thermodynamic conditions, an ice lattice comprising  10240 water molecules was rescaled to the corresponding bulk unit cell dimensions. Then, the system was elongated in the $z-$ direction so that the basal face (0001) was exposed to the vapor. Such prepared ice stack left half a bilayer on  each  side  of  the  slab to promote the premelting behavior. Afterwards, the sodium (calcium) and chloride ions have been randomly distributed on top of the ice interface and simulations were launched up to 600 ns in the canonical ensemble ($NVT$). The averages were collected from the last 100 ns of simulation time. Being aware of the limitations of the canonical ensemble for the description of binary mixtures (cf. Section~\ref{sec:thermo}), three surface coverages were examined for each temperature. Simulation box sizes for the different studied systems are shown in Table~\ref{tab:gamma}.  Examples of the initial configuration and an equilibrated sample are shown in Figure~\ref{fig:init} for NaCl deposited on the ice surface. 

Auxiliary bulk simulations have been performed for both pure water and electrolyte solutions which served as  a reference  for the properties of premelted layers formed at the ice-vapor interface.  

\subsection{Bulk transport properties}

Self-diffusion coefficients for both pure water and electrolyte solutions were calculated using the Einstein relation, involving the calculation of the mean-squared displacement (MSD) of individual water molecules, which can be cast as:

\begin{equation}
    \left<\Delta r^2(t)\right>=\left<(\mathbf{r}(t)-\mathbf{r}(0))^2\right>
    \label{eq:msd}
\end{equation}

\noindent where $\mathbf{r}(t)$ is the position of a water molecule at a time $t$, and the triangular brackets denote a thermal average over all time origins. 
The diffusion coefficient is then related to the slope of the MSD as $\left<\Delta r^2(t)\right>=6Dt$. 

The shear viscosity was calculated from the Green-Kubo formula:

\begin{equation}
    G_{\alpha\beta}=\frac{V}{k_BT}\left<j_{\alpha\beta}(t)j_{\alpha\beta}(0)\right >
\end{equation}

\noindent where $j_{\alpha\beta}$, with  $\alpha,\beta=x,y,z$ represent the components of the stress tensor, and $G_{\alpha\beta}(t)$ is the corresponding time correlation function.
The shear viscosity may be calculated exploiting all six independent elements of the correlation functions as: 
\begin{equation}
    \eta=\int_0^\infty G_{\eta}(t)dt
\end{equation}
\noindent where $G_{\eta}=\frac{1}{6}[G_{xx}+G_{yy}+G_{zz}
+\frac{3}{4}(G_{xy}+G_{xz}+G_{yz})]$ is a properly weighted linear combination of diagonal and off-diagonal elements \cite{daivis94}.

The simulation scheme involved two steps. In the first one, auxiliary $NpT$ simulations were performed for $10$~ns, allowing us to estimate the average box size and density of the fluid. Afterwards, the system was rescaled to the average dimensions estimated before and production runs lasting up to $30$~ns in the $NVT$ ensemble were launched to calculate both transport properties simultaneously from a single simulation. 

Having evaluated the self-diffusion coefficients and shear viscosity, we employed the Stokes-Einstein relation to calculate the hydrodynamic diameter at all considered state points. This can be cast as:

\begin{equation}
    a=\frac{k_BT}{3\pi\eta D}
    \label{eq:hydro-diameter}
\end{equation}

\noindent where $k_B$ is the Boltzmann constant.

\subsection{Freezing temperature depression}

To evaluate the freezing point depression of CaCl$_2$, the direct-coexistence method proposed by Noya \textit{et al.} was employed \cite{lamas22}. In this approach, a slab of aqueous electrolyte solution is brought into contact with bulk ice and the system is allowed to evolve, either by melting some ice (thereby decreasing the salt concentration) or by freezing some water (thereby increasing the salt concentration). According to the Gibbs phase rule, for a two-component system at fixed temperature and pressure, an equilibrium state corresponds to a two-phase coexistence at a specific salt concentration in the aqueous solution; this equilibrium state defines the liquidus line in the bulk salt-water phase diagram. Initial concentrations of an aqueous solution were chosen so that they were close enough to the experimental values \cite{oakes90} to facilitate reaching equilibrium value. They were equal to $c=1.$ mol/kg for $T=260$~K, $T=262$~K, and $T=264$~K and $c=2.$ mol/kg for $T=250$~K and $T=255$~K. For these simulations, we exposed the pII ice interface since it exhibits the fastest dynamics \cite{nada05}.  Runs were launched in $Np_\perp AT$ ensemble for $1~\mu s$, followed by $0.5~\mu s$ to calculate the average concentration in the middle region of the brine. 

\subsection{Analysis of the premelting layer}\label{subsec:prem}

To identify molecules in different environments, the CHILL+ order parameter was used \cite{nguyen15}. This allows to distinguish between ice Ih, ice Ic, clathrates, interfacial ice Ih, and liquid-like phases. The number of solid-like and liquid-like atoms was determined following the same procedure as in our recent work \cite{baran22}. First, the largest ice cluster was identified by lumping ice Ih, interfacial ice Ih and ice Ic into the solid-like category, leaving  the remaining water molecules classified as liquid-like. With water molecules labeled according to these criteria at each time frame of the trajectory, it is possible to define an effective molal concentration of the brine atop the ice surface as: 

\begin{equation}
    c=\frac{N_{\text{ion}}}{N_{\text{liq}}\cdot M_w}
    \label{eq:conc}
\end{equation}

\noindent where $N_{\text{ion}}$ is the number of ions at an interface (c.f.  Table~\ref{tab:gamma}), $N_{\text{liq}}$ is the number of liquid-like water molecules, and $M_w=0.018$~kg/mol. Of course, a concentration in an inhomogeneous system is a function of position, and is not strictly well defined for a premelting film. In practice, the ions are very strongly segregated within the liquid layer, and we find that this measure of effective concentration is quite useful and serves as a measure for comparison with bulk properties.

Next, the ice-fluid (i-f) and liquid-vapor (l-v) interfaces were located using a local geometrical criteria.  
To identify the i-f surface, all solid atoms lying within a rectangular area of size $a\times b$, corresponding to the ice unit cell dimensions at the given thermodynamic conditions and centered at $\mathbf{r}$, were found. The surface height at $\mathbf{r}$, $z_{if}(\mathbf{r})$, is then determined from the average position of the four topmost (bottommost) solid atoms of the upper (lower) interface of the largest solid cluster. At the same point $\mathbf{r}$, the l-v interface height $z_{fv}(\mathbf{r})$ is located by averaging four topmost (bottommost) liquid-like atoms of the upper (lower) interface lying within a square area of $3\sigma\times3\sigma$. The surfaces are determined over a grid on the plane of the interface. The grid has 16 and 4 points in the $x$ and $y$ direction, respectively. The film height of a premelting layer was then calculated as
an average over all local film thicknesses:

\begin{equation}
    h=\frac{1}{N}\sum_i^N \lbrace z_{fv}(\mathbf{r}_i)-z_{if}(\mathbf{r}_i) \rbrace
    \label{eq:premelting}
\end{equation}

\noindent with $N$ the total number of grid points on the surface.




Finally, transport properties of liquid-like molecules within the previously identified interfacial region were calculated. First, we monitored the mean squared displacement (Eq.~\ref{eq:msd}) of water-like molecules in the parallel direction to the surface. The parallel diffusion coefficient in the premelting layer was then   related to the slope of MSD as $<\Delta r^2(t)>=4D_\parallel t$. Ultimately, the shear viscosity of the interfacial layers was estimated from the Stokes-Einstein relation as:

\begin{equation}
    \eta=\frac{k_BT}{3\pi D_\parallel a}
    \label{eq:se-visc}
\end{equation}

\noindent where $a$ is the hydrodynamic diameter estimated from properties of the bulk electrolyte solution at the same temperature and concentration as the quasi-liquid layer.  

\subsection{Simulation details}

Molecular dynamics simulations were launched using the LAMMPS simulation package \cite{lammps22}. Trajectories were evolved using the velocity-Verlet integrator with a time step of 1 fs, employing a quaternion-based rigid-body algorithm \cite{kamberaj05}. The temperature was controlled using a Nos{\'e}-Hoover chains algorithm with a damping factor of $\tau=1$~ps. Tail corrections were not included. Long-range electrostatics were computed using the particle–particle particle-mesh method \cite{darden93}, with charge structure factors evaluated using a fourth-order interpolation scheme and a grid spacing of 1 \AA.

\section{Results and discussion}

Knowing that the melting point of the TIP4P/Ice water model is approximately $T_m\approx270$~K \cite{conde17} and that the addition of ions causes the freezing point depression, temperatures from $T=265$~K down to $T=250$~K were investigated.  The lowest temperature studied is close to the experimental eutectic point of the H$_2$O-NaCl binary mixture,  which lies at $T\approx -21$~$^{\circ}$C. For each temperature, three different salt coverages were studied, $\sigma=0.55,\;1.1$ and $2.2$~salt molecules per nm$^2$.   

To complement our study, we also 
studied premelting films formed by addition of calcium chloride at equal temperatures and surface coverages. To put the results in context, the liquidus line of the Madrid force field for this salt was also calculated. 

\subsection{Structure}

\begin{figure}
    \centering
    \includegraphics[width=\linewidth]{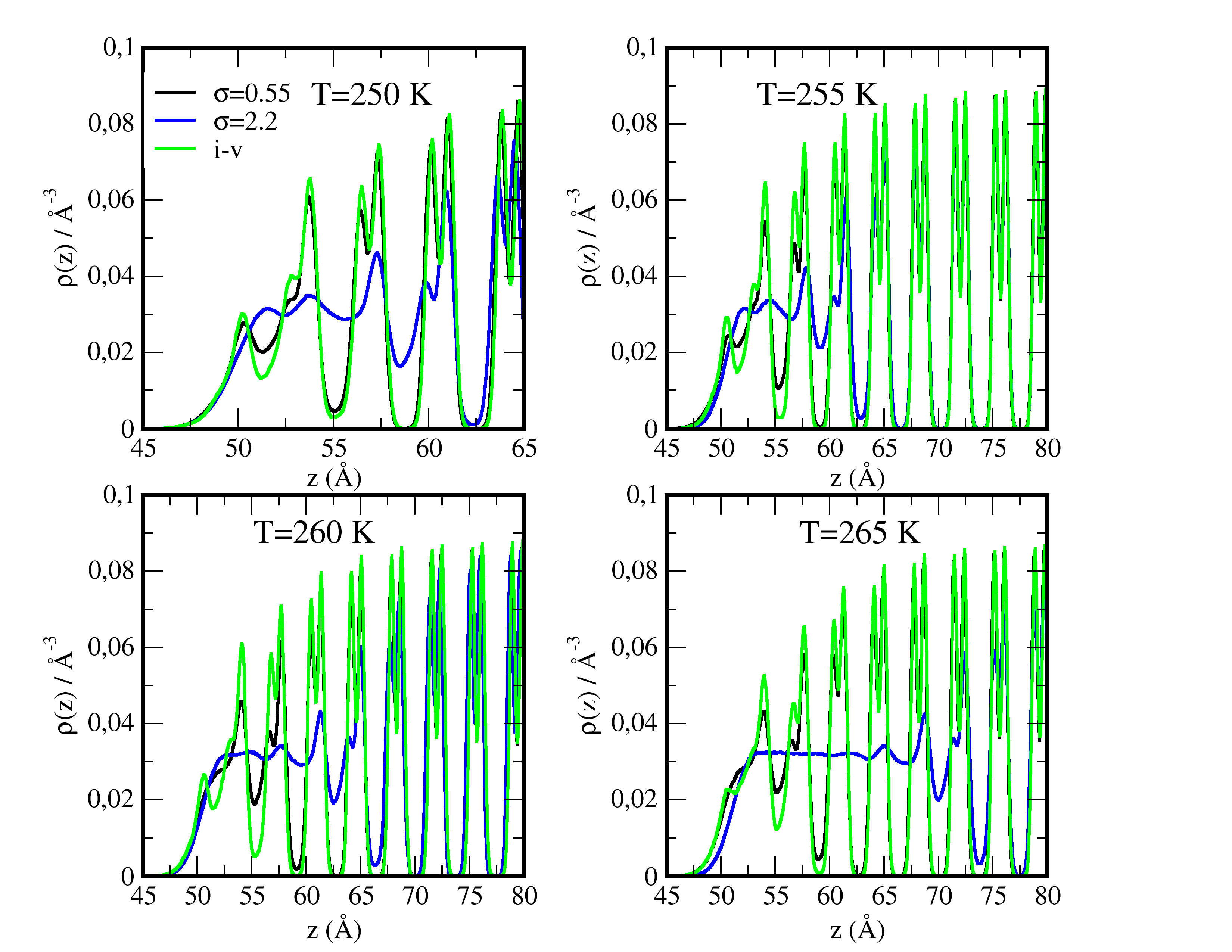}
    \caption{Premelting structure of ice with adsorbed NaCl. Figure shows the number density profiles $\rho(z)$ at four different temperatures. Results for films with $\sigma=0.55$ and $\sigma=2.2$ are compared with those of pure ice premelting. }
    \label{fig:dens-prof}
\end{figure}

\begin{figure}
    \centering
    \includegraphics[width=\linewidth]{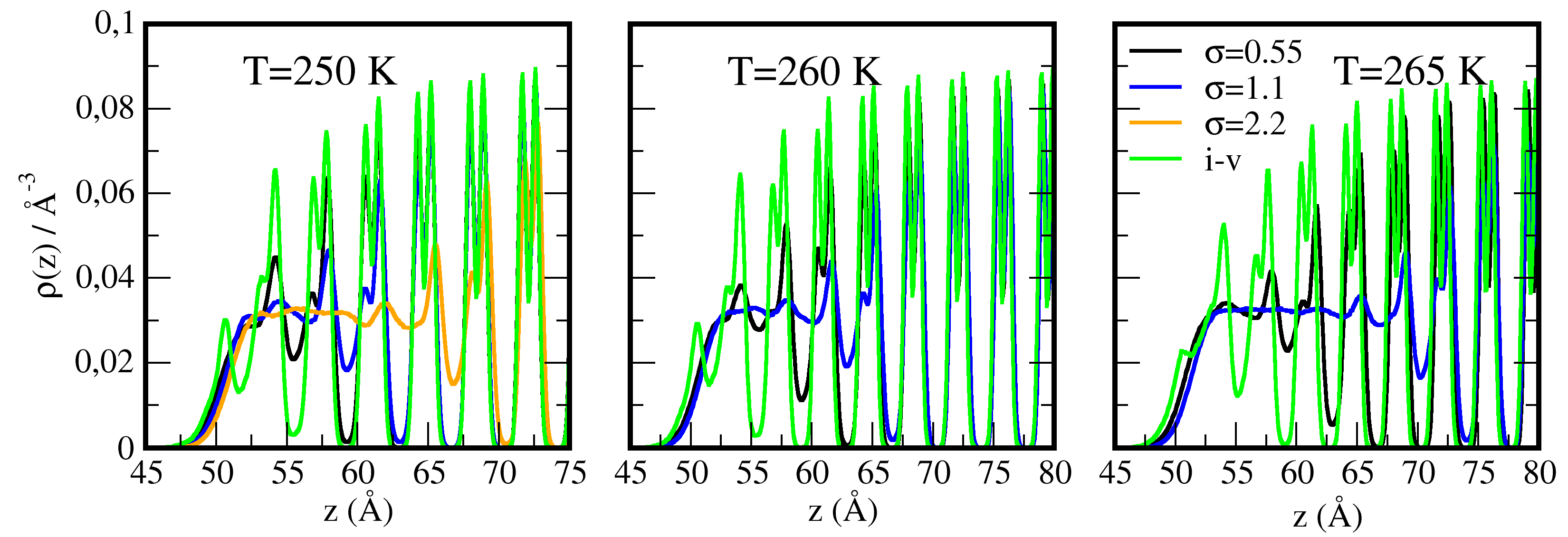}
    \caption{
    Premelting structure of ice with adsorbed CaCl$_2$. Figure shows the number density profiles $\rho(z)$ at three different temperatures. Results for films with surface coverage $\sigma=0.55$, $\sigma=1.1$ and  $\sigma=2.2$ are compared with those of pure ice premelting.
    }
    \label{fig:dens-prof-cacl2}
\end{figure}

\begin{figure}
    \centering
    \includegraphics[width=\linewidth]{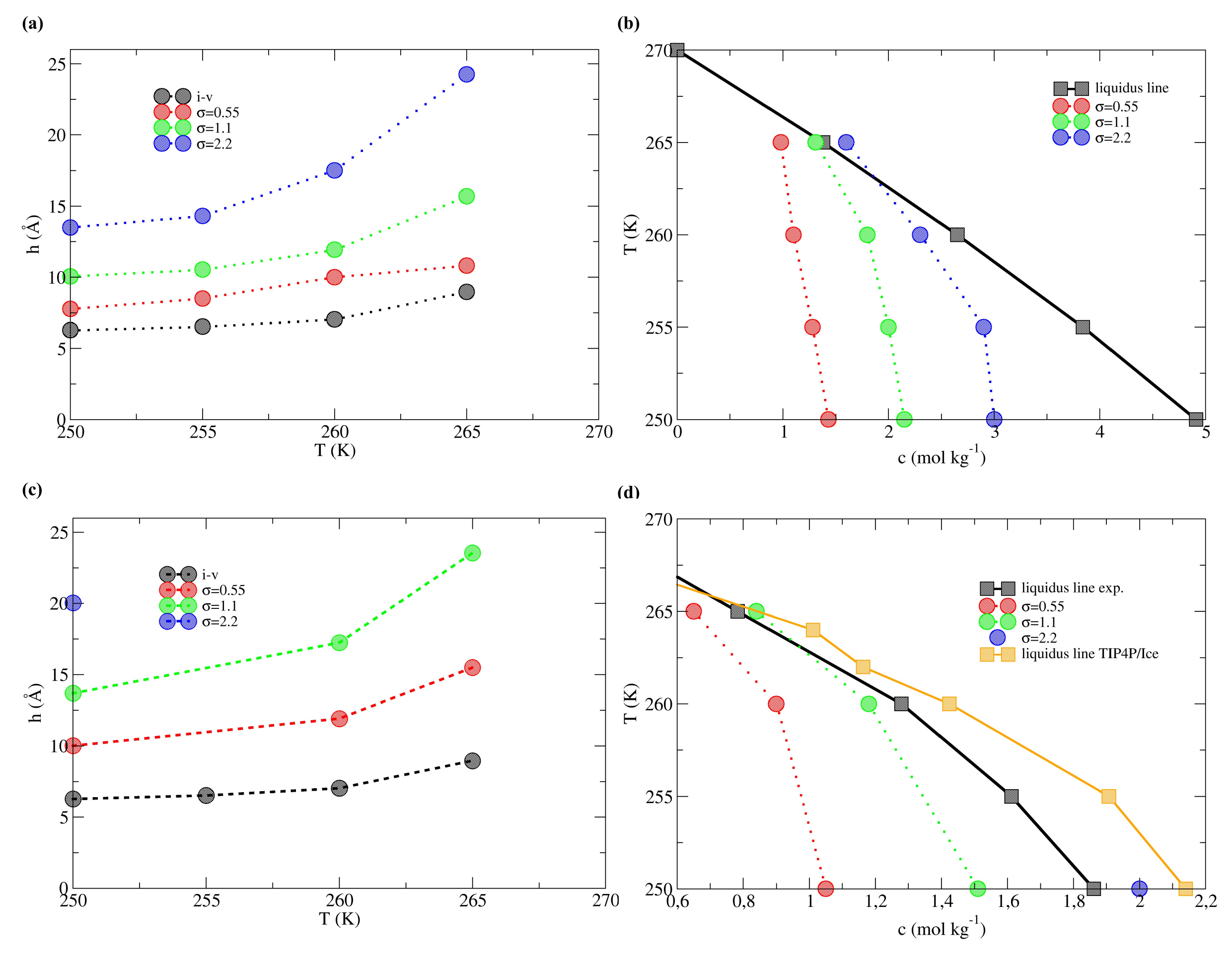}
    \caption{Premelting film heights as a function of temperature for the ice-vapor (i-v) reference system and three examined surface coverages for NaCl (a) and CaCl$_2$ (c). Equilibrium concentrations formed at the ice-vapor interface with a liquidus line from the bulk phase diagram for NaCl (b) and CaCl$_2$ (d). 
    The reference black solid line is taken from Ref.~\cite{blazquez24} for NaCl and  from Ref.\cite{oakes90} for CaCl$_2$.} 
    \label{fig:premelting}
\end{figure}



The discussion begins with an analysis of the premelting film's structure.  Figure~\ref{fig:dens-prof} presents the density profiles for the lowest and highest surface coverages of NaCl and their comparison with the pure ice-vapor system. The signature of premelting is the loss of the bilayer structure on the density profile at the ice-vapor interface, typical for the basal facet of bulk ice. It is apparent that the extent of premelting is larger for briny ice than for the ice-vapor interface in the absence of ions. For a given ion's surface coverage, $\sigma$, the same trend can be observed as for neat ice - higher temperatures enhance the premelting. In analogy, for a fixed temperature the premelting increases with increasing surface coverage, as observed previously \cite{berrens22}.  Particularly, at the highest surface coverage $\sigma=2.2$, the extent of premelting  reaches a thickness of more than 2~nm at $T=265$~K, and almost a plateau liquid like density is achieved within the center of the film.  Another interesting feature revealing the approach to bulk liquid behavior is that a characteristic "bump" at the ice-vapor interface for pure-water conditions is smoothened upon addition of ions.

Essentially the same conclusions as for NaCl adsorption can be drawn from the analysis of the premelting film structure when CaCl$_2$ is deposited on the ice-vapor interface. Figure~\ref{fig:dens-prof-cacl2} shows the density profiles for different surface coverages. As in the NaCl case, the extent of premelting is larger for briny ice than for the pristine ice-vapor interface; however, it is even more pronounced for CaCl$_2$ than for NaCl. Notably, the highest surface concentration $\sigma=2.2$, is shown only at the lowest temperature ($T=250$~K) because at higher temperatures the entire ice stack has melted. This implies that an even larger ice stack would be required to reliably estimate the premelting film thickness under these conditions. This suggests that  premelting films with a CaCl$_2$ coverage above $\sigma=2.2$ become surface melted at temperatures above 255~K.

To confirm the formation of premelting layers, the CHILL+ algorithm was invoked \cite{nguyen15} to identify molecules pertaining to the solid-like and liquid-like environments (cf. Section~\ref{subsec:prem}). Afterward, the premelting film heights, $h$, were calculated using Eq.~\ref{eq:premelting}. Figure~\ref{fig:premelting}-a,c present the relation of film thickness with temperature for NaCl and CaCl$_2$. It is clear that for all considered temperatures and surface coverages, the premelting film is thicker than in the pure-water reference systems, consistent with the density profiles. Moreover, the film heights for CaCl$_2$ are much thicker as compared to the NaCl case.

To gain deeper insight into the nature of the binary premelting films, the concentration of the interfacial layer was calculated using Eq.~\ref{eq:conc}, with the assumption that ions do not move into the vapor phase and they do not penetrate into the ice lattice. The validity of this assumption is assessed later. The results are displayed in Figure~\ref{fig:premelting}-b) and d) for NaCl and CaCl$_2$, respectively. 

Focusing first in the sodium chloride case at low temperature (Figure~\ref{fig:premelting}-b), we observe, as expected from the discussion in section II that the film concentrations for the thin premelting films observed at T$=250$~K are smaller than that at the liquidus point in the range of surface coverages studied. Increasing the temperature at fixed coverage appears to decrease the film's concentration almost linearly. However, on approaching the liquidus line, the film concentrations bend strongly and appear to approach the liquidus line almost tangentially.

Regarding the CaCl$_2$ case, the liquidus line of this system for the Madrid force field had not yet been calculated. The results from our simulations, as shown with orange symbols in Figure Fig.~\ref{fig:premelting}-d) are compared with experimental results. The comparison shows that the Madrid force field is quite satisfactory down to about 260~K. At lower temperatures it predicts somewhat too large liquidus concentrations, but it is able to show strong deviations from ideality as is the case in experiment (deviations from ideality being revealed as deviations from linear behavior of the liquidus line).

Be as it may, the results found for the film's concentrations follow the same trend observed for premelting films with NaCl. The film concentrations decrease as temperature increases, and bend in order to meet in close to tangential fashion the liquidus line. 
The main difference with NaCl is that the liquidus line of CaCl$_2$ lies at significantly smaller concentrations. Therefore, the fixed surface coverages chosen become relatively larger. As a result, films with intermediate coverage are already very close to the liquidus line at 260~K, and have fully met the liquidus line at 265~K. For the highest coverage, the film concentration is already close to the liquidus at the lowest temperature, while for the other two temperatures, the outcome of the simulations was the full melting of our samples. i.e., we were unable to stabilize the premelting films of highest coverage at near triple point conditions. A larger system would be required to further explore whether such films can be stabilized, but this unfortunately poses a considerable computational burden. 

Our analysis based on a nominal premelting film concentration appears to be a very useful tool to identify genuine surface phenomena from possible bulk three phase behavior. The results observed in Figure~\ref{fig:premelting}-b and Figure~\ref{fig:premelting}-d appear to confirm the expectations discussed in Section~\ref{sec:thermo}, namely, that the liquidus line of the bulk two phase system serves as a reference which bounds from above the maximum brine concentration in the premelting film.

Notice, however, that both for the NaCl and CaCl$_2$ samples there are  a few surface states which have very nearly achieved the equilibrium liquidus concentration.

Particularly, at $T=265$~K, the NaCl films with the two highest surface coverages, and the CaCl$_2$ film with intermediate concentration have attained the liquidus concentration. In this situation, it must be concluded that the bulk thermodynamic fields determining the system's thermodynamic state must be at or exceedingly close to the triple point conditions. Two possible equilibrium outcomes could be envisaged in this situation in an open system. Either the film remains at its current finite thickness, which would imply that the ice surface is not wet by the brine; or the film thickness grows boundless, corresponding to the brine wetting the ice surface fully. Unfortunately, to decide upon these two possible outcomes, we would need to perform simulations in a grand canonical system, where arbitrary amounts of either salt or water are allowed so that the system choses the thermodynamic state without finite system size constraints. In the closed   system studied here, the observation is that indeed  thick premelting films of either NaCl or CaCl$_2$ brines remain equilibrated in presence of bulk ice and vapor phases at 265~K, suggesting that at least finite metastable premelting films can be formed on the ice surface in very nearly triple point thermodynamic conditions.

Concerning the dynamics of film growth formation, it is worth mentioning that the process of reaching an equilibrium concentration of the premelting film can take as long as half a microsecond of the simulation time for NaCl (cf. Figure S1) or even more for CaCl$_2$ (cf. Figure S2). This can be attributed to two main reasons. The first one is the competition between two opposite effects -  the ions promote melting so that they can reach the bulk equilibrium concentration but simultaneously, water molecules conspire in the opposite direction to prevent premelting thickness grow beyond the equilibrium premelting thickness of pure water.  The second effect is model-specific. Madrid-2019 force field underestimates the diffusion of ions, despite a scaled value of the ionic charge. Although it is a quite robust model, it has been reported that different scaling factors need to be considered depending on whether one seeks for accurate predictions of bulk thermodynamic properties, surface tensions, or transport properties \cite{blazquez23, breton20}. This artifact of the model is likely to slow down the overall melting dynamics.

\begin{figure}
    \centering
    \includegraphics[width=\linewidth]{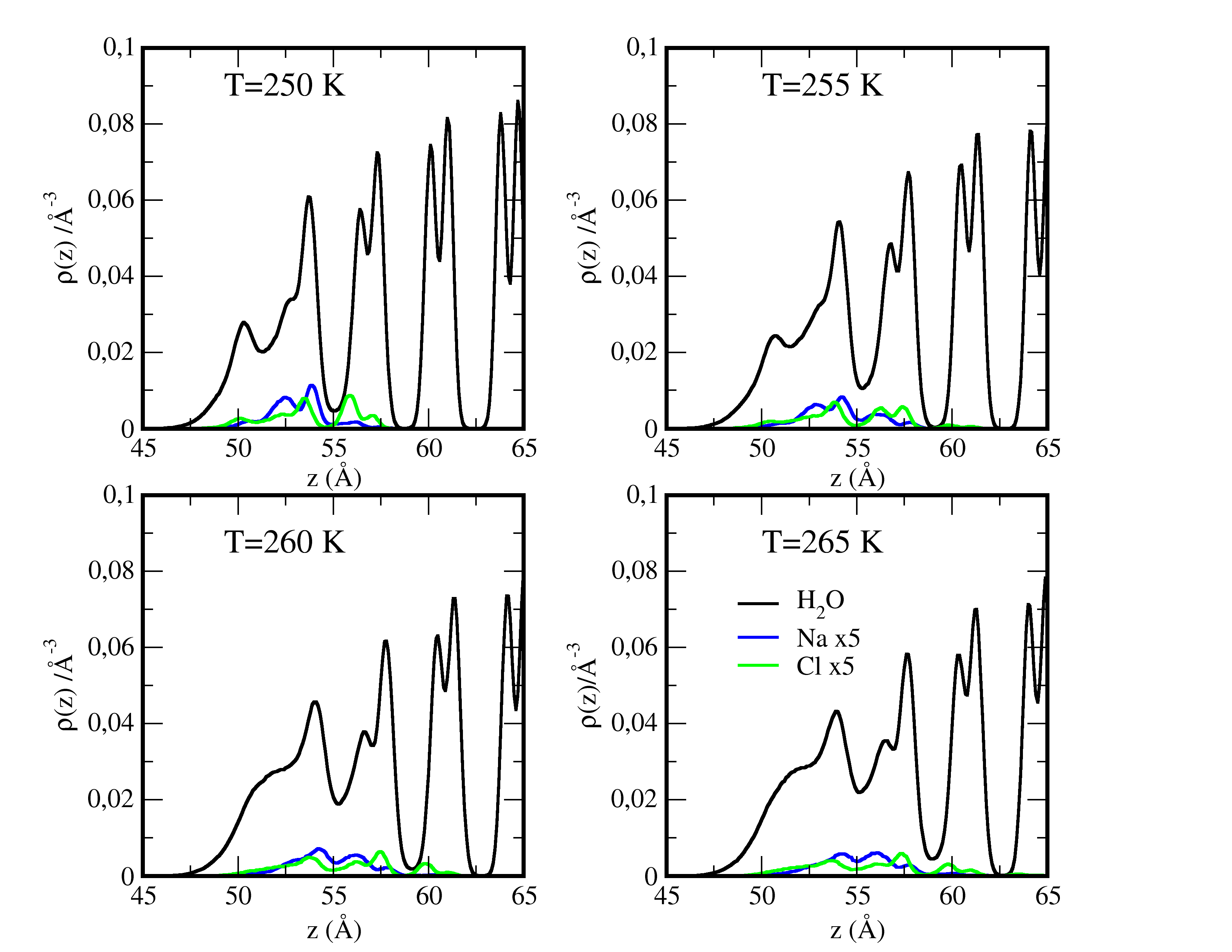}
    \caption{The comparison of number density profiles $\rho(z)$ of ice adsorbed with NaCl, calculated for each component separately at four different temperatures. Results for surface coverage equal to $\sigma=0.55$. The scale for the ions is magnified 5 times for better visualization.}
    \label{fig:dens-ions-055}
\end{figure}

\begin{figure}
    \centering
    \includegraphics[width=\linewidth]{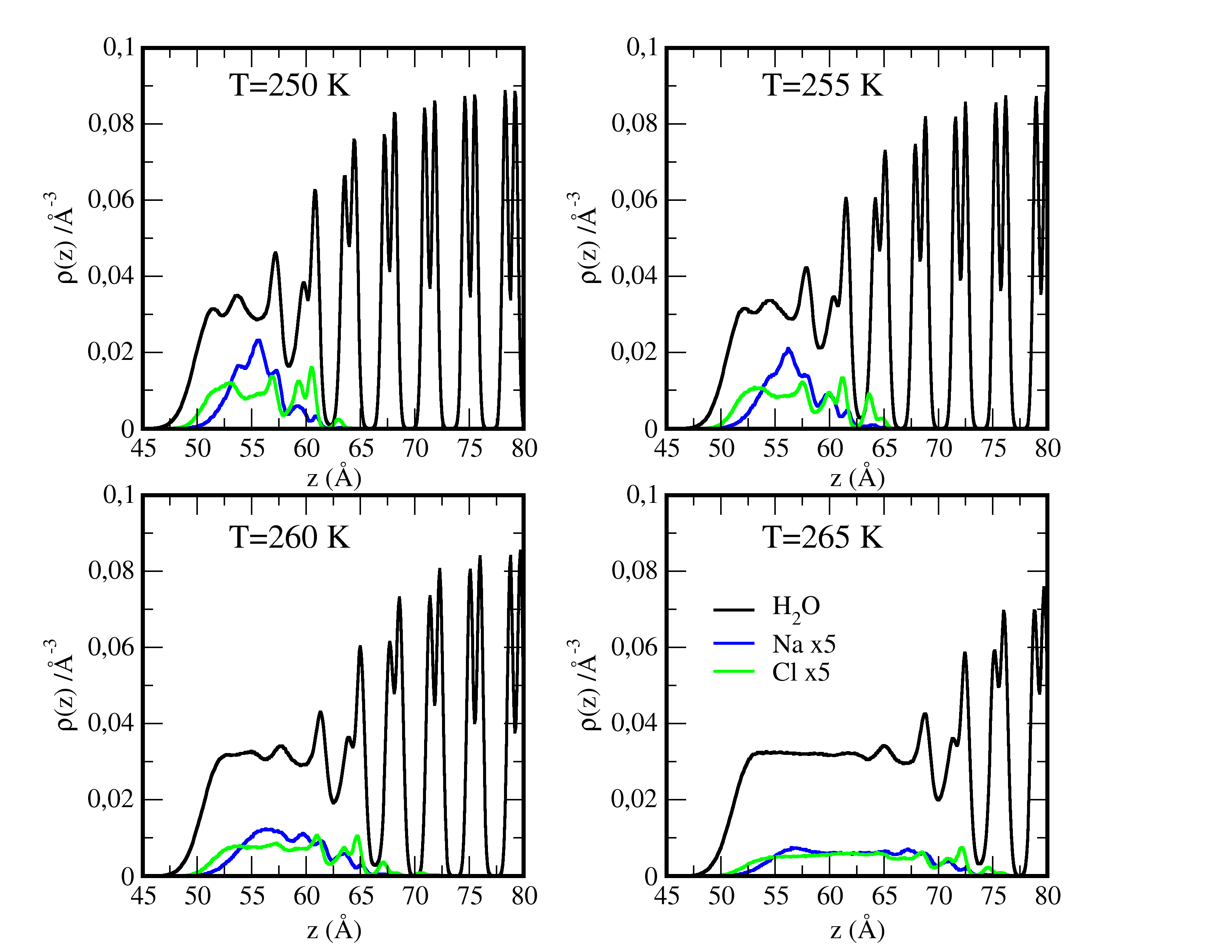}
    \caption{The comparison of number density profiles $\rho(z)$ of ice adsorbed with NaCl, calculated for each component separately at four different temperatures. Results for surface coverage equal to $\sigma=2.2$. The scale for the ions is magnified 5 times for better visualization.}
    \label{fig:dens-ions-217}
\end{figure}

\begin{figure}
    \centering
    \includegraphics[width=\linewidth]{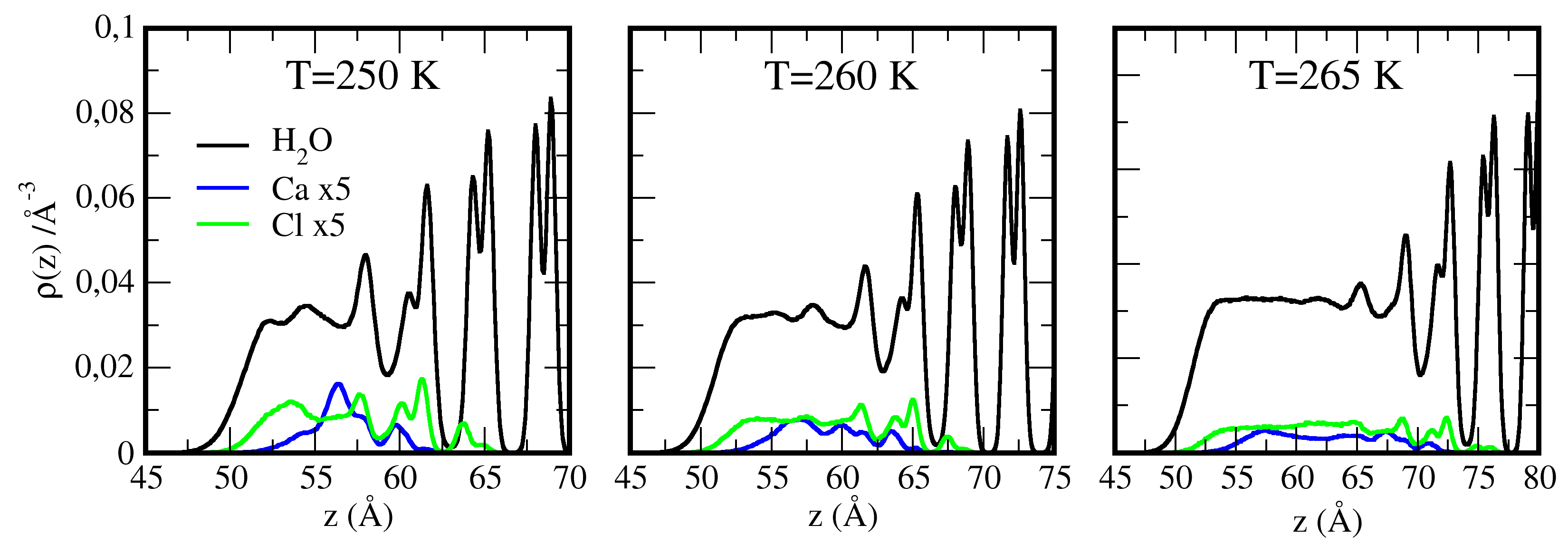}
    \caption{The comparison of number density profiles $\rho(z)$ of ice with adsorbed CaCl$_2$, calculated for each component separately at three different temperatures. Results for surface coverage equal to $\sigma=1.1$. The scale for the ions is magnified 5 times for better visualization.}
    \label{fig:dens-ions-110-cacl2}
\end{figure}

In order to understand how the ions are arranged within the premelting film, 
 the density profiles of individual components were calculated. The results are presented in Figures.~\ref{fig:dens-ions-055} and ~\ref{fig:dens-ions-217} for sodium chloride and in Figure~\ref{fig:dens-ions-110-cacl2} for calcium chloride. Ion stratification at the surface is readily observed for both salts at all surface coverages, although it is much better pronounced for $\sigma=2.2$. 
 The figure shows that both the liquid-vapor and ice-liquid surfaces bounding the film are enriched with the anions, while the cations are depleted at the interface and concentrate within the center of the film.  


Interestingly, the degree of ionic stratification increases as the extent of premelting decreases, as evidenced by the highly structured films at $T=250$~K and $T=255$~K for $\sigma=2.2$, and at $T=250$~K for $\sigma=1.1$. In contrast, at $T=260$~K and $T= 265$~K the premelting layer in the presence of both salts becomes sufficiently thick that the ions are distributed nearly uniformly throughout the film. In this limit, the density profiles show a smooth decay at the liquid-vapor surface, which has a clear preference for the anions.  This tendency of the anions to be distributed towards the vapor phase has been observed for brine-vapor interfaces of several salts \cite{litman24, jungwirth06}. Therefore, we find that the structural organization of ions at the brine-vapor interface of bulk solutions serves as a good proxy for ion organization in premelting films. 

At the opposite side of the premelting film,
another notable feature is the tendency for the anions to penetrate the ice lattice. It is worth noting that at some cases, the chloride anions can be found as deep as in the second bilayer of the bulk ice phase. According to M. Conde \textit{et al.} \cite{conde17b}, chloride anions can substitute two water molecules in the lattice nodes without disrupting the crystal structure. In contrast, sodium cations are never found to diffuse into an ice lattice in our simulations. As explained by M. Conde \textit{et al.}, this behavior is attributed to energetically unfavorable occupation of sodium cations in the interstitial sites that create local disruptions of the electroneutrality of the ice lattice. In similar fashion, calcium cations are never found to migrate into the bulk crystal. This results appear consistent with  the experimental observation that anions are usually more likely to penetrate deep into the ice lattice \cite{sivells23}.

The fact that the  anions  enrich the solid-liquid interface and penetrate into the first solid bilayer means that the effective concentrations of a brine formed in a premelting layer, shown in Figure~\ref{fig:premelting}-b, are subject to some error. Nevertheless, this discrepancy is negligible as long as the number of ions incorporated into the  lattice is much smaller than the total number of ions distributed in the entire interfacial region, which is the actual situation in our simulations.

\subsection{Dynamic properties}

\begin{figure}[h!]
    \centering
    \includegraphics[width=\linewidth]{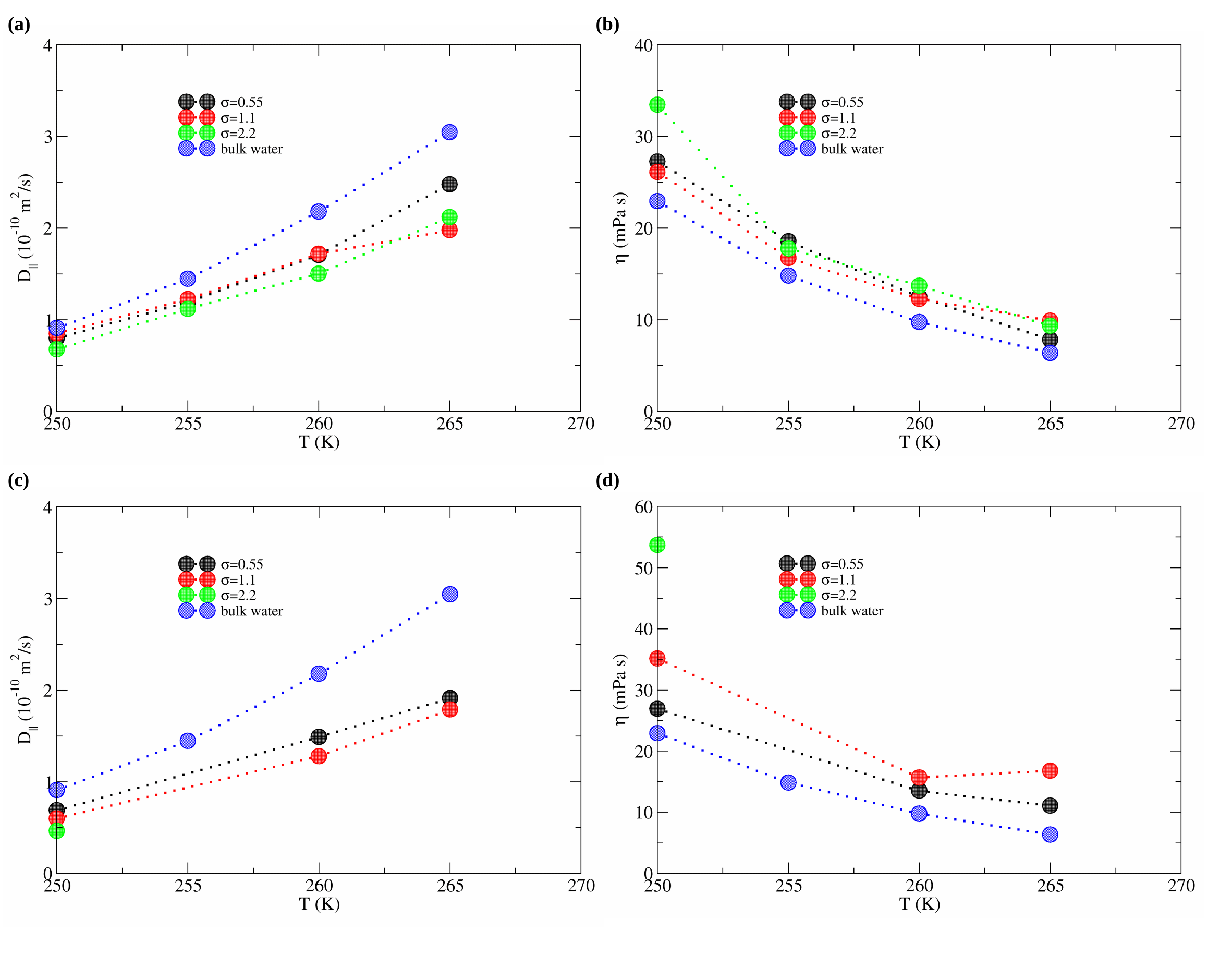}
    \caption{Transport properties of premelting layer with adsorbed salts. 
     Parallel self-diffusion coefficient (a, c) and shear viscosity (b, d) of the premelting layer in the presence of NaCl (a, b) and CaCl$_2$ (c, d) are compared with bulk pure water. 
    }
    \label{fig:transport}
\end{figure}

To complement the analysis of the premelting films, the rheological properties were examined. The results are shown in Figure~\ref{fig:transport}. The parallel self-diffusion coefficients of water molecules within the premelting films in the presence of sodium chloride are found to be qualitatively similar to those obtained from simulations of pure supercooled  bulk water. In contrast to NaCl solutions, the self-diffusion for premelting films with CaCl$_2$ are almost two times smaller at higher temperatures. 

Salts can be classified as \textit{structure makers} or \textit{structure breakers} based on their effect on the mobility of water molecules in aqueous solutions. The former suppress water mobility as concentration increases, whereas structure breakers initially enhance water self-diffusion upon salt addition, reaching a maximum before diffusion decreases at higher concentrations. NaCl and CaCl$_2$ are typical examples of structure makers, however, the distinction is not clear and it depends on the thermodynamic conditions. In particular, a crossover from structure-making to structure-breaking behavior has been reported upon cooling for several salts \cite{kim08,sedano24}.

Consistent with their structure making behavior, diffusion coefficients of water molecules in premelting layer containing adsorbed NaCl and CaCl$_2$ are smaller than in bulk water, with the difference decreasing as the temperature decreases. In addition, CaCl$_2$ forms stronger solvation shells at higher temperatures which more effectively suppress the mobility of water molecules. In contrast, for sodium chloride at $T=250$~K the self-diffusion approaches that of pure water, indicating that NaCl disrupts the hydrogen bond network formation, making the liquid solution less structured. 
Such a crossover from the structure-making to structure-breaking regime has been already reported for supercooled sodium chloride solutions \cite{garbacz14,sedano24}. No such crossover is observed for calcium chloride in the range of temperatures studied here. These results suggest that the rheology of the premelting film reflects trends observed in the aqueous electrolyte solutions. 

By virtue of the Stokes-Einstein relation, one would expect  the shear viscosity of premelting films to exceed that of bulk water. Using hydrodynamic diameters of bulk aqueous NaCl and CaCl$_2$ solutions at concentrations corresponding to those in the premelting films, together with the parallel diffusion coefficients, the shear viscosities were estimated via Eq.~\ref{eq:se-visc}. The results shown in Figure~\ref{fig:transport}-b,d confirm that the viscosities are consistently higher than those of bulk water, in agreement with expectations. Moreover, at comparable temperatures and surface coverages, premelting films containing CaCl$_2$ systematically exhibit higher viscosities than those containing NaCl. This behavior is consistent with the stronger hydration of the divalent cation which enhances local structuring and suppresses water molecular mobility.  

\begin{figure}[h!]
    \centering
    \includegraphics[width=\linewidth]{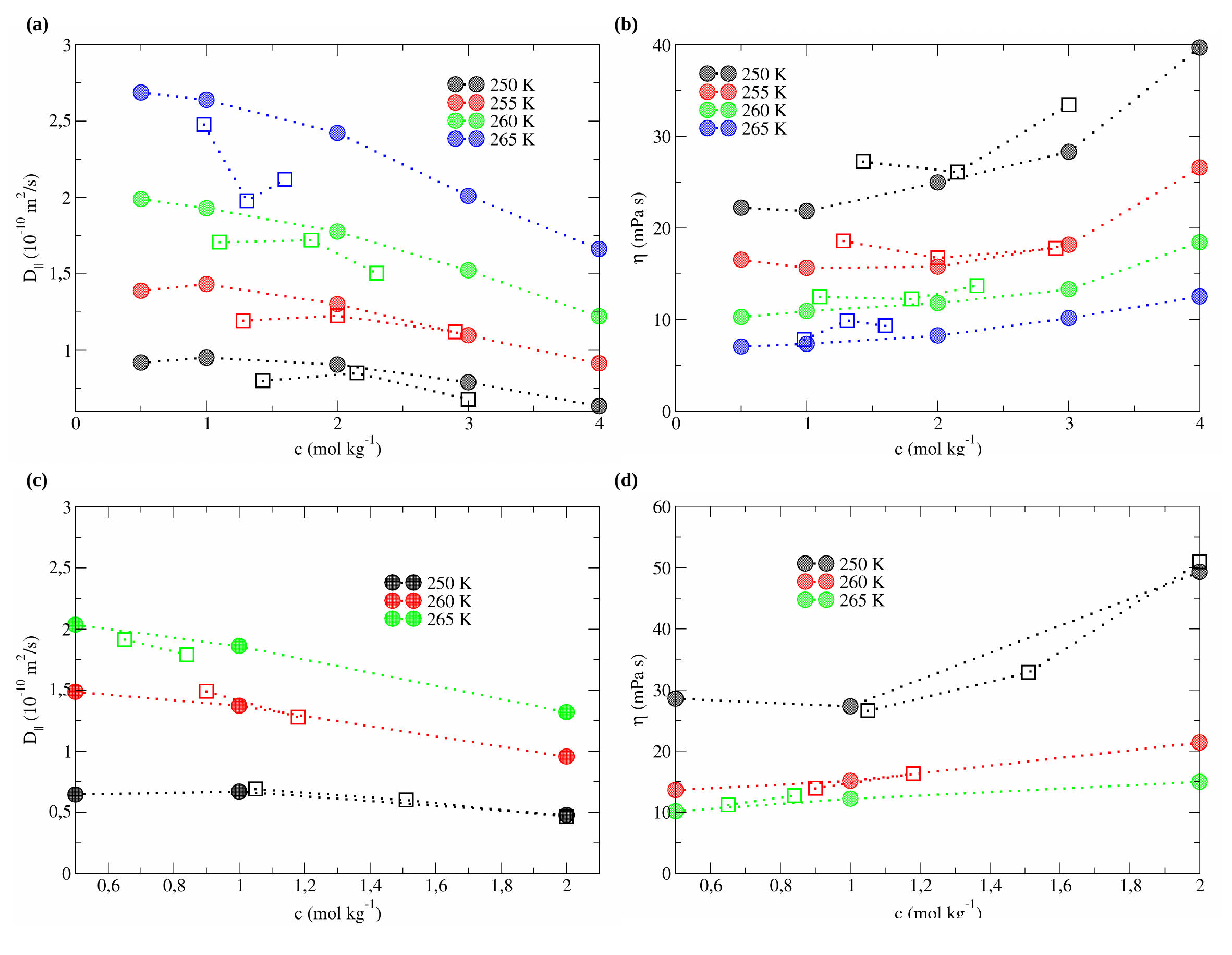}
    \caption{Transport properties of premelting layer with adsorbed salts. 
     Parallel self-diffusion coefficient (a, c) and shear viscosity (b, d) of the premelting layer (filled squares) in the presence of NaCl (a, b) and CaCl$_2$ (c, d) are compared with the corresponding bulk aqueous solutions (open circles). 
    }
    \label{fig:transport-brine}
\end{figure}

The deviations in rheological properties from pure-water conditions that are observed for the premelting films in the presence of ions suggest that the comparison should be better made with bulk electrolyte solutions at the same temperature and concentration. The results, shown in Fig.~\ref{fig:transport-brine} demonstrate excellent agreement for both self-diffusion coefficients and shear  when compared this way. Notably, the rheological properties of calcium chloride are nearly identical to their bulk counterparts, whereas sodium chloride exhibits slight deviations. 

This difference can be attributed to the substantially larger premelting film thicknesses in the calcium chloride case, which more closely approach bulk-like conditions. In contrast, the thinner films observed in the presence of sodium chloride remain more strongly influenced by the ice-water interface, where surface interactions modify molecular mobility. This effect is manifested by a more pronounced stratification of ions within the premelting layer. Overall, these results highlight that  bulk electrolyte solutions provide an appropriate reference state for estimating the rheological properties of premelting films in the presence of salts. 

The similarity of rheological properties of premelting films with the corresponding bulk properties was also observed for interfacially premelting films of pure water \cite{baran22}. There are two crucial issue for the success of this comparison
that need to be highlighted, as opposite results have often been reported. Firstly, the comparison must be made for the parallel components of the diffusion coefficient. Indeed, the presence of an inhomogeneity in directions perpendicular to the film means that the diffusion along this direction is severely affected by the confinement, and cannot be equated to the corresponding expectation for the unconfined liquid. Secondly, the calculation of the mean squared displacement in the parallel direction must be restricted to  water molecules that remain liquid-like, as labeled by a reliable order parameter. Such restriction does not prevent the reliable calculation of the mean squared displacement, as the typical residence time of liquid-like molecules within the premelting film is much larger than that required for molecules to  acquire diffusive behavior. With this caution, our results show that the bulk liquid serves as a reliable proxy for the qualitative description of the film's rheological properties.

\section{Conclusions}


Understanding the salt uptake on the surface of briny ice is an important problem of geophysics, with applications in glacier motion, frost heaving and gas scavenging \cite{dash06}. Despite this interest, only a very limited number of studies have dealt with this problem before \cite{berrens22,hundait17,ribeiro21}. A major difficulty is that it is very difficult to control and monitor the bulk thermodynamic fields that are necessary to set the nearly three phase equilibrium conditions of the system \cite{wettlaufer99}. This leads to a very difficult characterization of the premelting films, as it is usually not  possible to distinguish surface phenomena from three phase coexistence \cite{cho02}.

In this study, we have performed a rigorous thermodynamic analysis of the conditions of equilibrium premelting in binary mixtures of ice and salt. Particularly, by introducing a geometrical criteria for the film thickness, and an effective premelting 
layer concentration, we are able to identify genuine surface layers from three bulk phase coexistence, and relate the properties of the films to those of  bulk equilibrium brine solutions.

Our method of analysis is applied to study  the adsorption of sodium and calcium chloride on the ice surface. The results confirm that these salts significantly enhance ice premelting, and lead to the formation of quasi-liquid briny solutions that can be twice as thick as those found on the surface of pure ice. Particularly, for the moderate to large surface coverages studied in our work, we find premelting films that range from one nanometer to more than two nanometers in thickness. Despite this nanoscale size, we have shown that the dynamic properties of the films are already quite close to those found in bulk solutions of equal concentration.

Thanks to our method of analysis, we are able to show for the first time that the briny premelting layers  exhibit effective concentrations that lie well within the immiscibility gap of the ice-brine equilibrium phase diagram. As the surface coverage gradually increases, however, the layer concentrations  smoothly converge to the equilibrium bulk liquidus line. Our results clarify for the first time that surface layers of finite extent can be equilibrated as the two phase ice-vapor system reaches the three phase ice-brine-vapor triple point.




As far as the structure of the premelting films is concerned, we find that  the ions in the premelting film are strongly  stratified, with the chloride anions exhibiting a preference to adsorb next to the interfaces, and the cations remaining preferentially at the center of the layer.  This mechanism is the more pronounced the thinner the premelting film is, thus at lower temperatures. In contrast, at higher temperatures and larger surface coverages, the premelting film becomes sufficiently thick that the ionic distribution approaches a plateau value, although anions remain preferentially located near the interface. At the liquid-vapor side of the premelting layer, we find the films closely resemble those formed at aqueous electrolyte-air interfaces. 
On the other hand, at the opposite brine-ice  interface, the anions can also be engulfed into the ice lattice. This engulfment takes place by the replacement of two water molecules by one chlorine ion, with minor disruption of the crystalline structure \cite{conde17b}.  

These results suggest that the commonly invoked assumption that ions enhance premelting film thickness holds for salts that do not interact strongly with an ice phase and only at the temperatures above the eutectic point of the relevant salt-water phase diagram. Below this temperature, salt precipitation is expected to occur, and the influence of ions on the premelting film thickness cannot be inferred from the present simulations. Regardless of this, sodium chloride appears as a suitable proxy to describe seawater ice not only because it constitutes approximately 85\% of seawater composition, but also because its eutectic point lies approximately $20^\circ$~C below the melting point of pure water, placing it well in the environmentally-relevant conditions.  The only relevant salts at this conditions could be potassium and magnesium chlorides but their amount is negligible and likely would not affect the conclusions drawn here. In contrast, sulfates that are also present in seawater exhibit eutectic points that are at most a few Kelvins below the melting point, so it is expected that they will precipitate out of the premelting solution and contribute little to the observed properties.

In summary, we have shown that using a suitable geometrical definition of premelting film thickness, it is possible to meaningfully measure the effective salt concentration of briny quasi-liquid layers. This allows to relate the properties of  the premelting layers to bulk solution properties, and provides   a detailed characterization that was hitherto not possible \cite{berrens22,hundait17,ribeiro21}. We expect that the improved understanding of briny premelting films will  find applications ranging from the study of premelting layers formed   from frozen sea sprays, to the sliding of glaciers and skating on  sea water ice.

\section*{Acknowledgment}
LGM  would like to thank financial support from the Agencia Estatal de Investigaci\'on (Ministerio de Ciencia y Econom{\'i}a) under grant PID2023-151751NB-I00. \L B would like to thank the Polish National Agency for Academic Exchange, under Grant No. BPN/BEK/2023/1/00006 Bekker 2023.  We also benefited from generous allocation of computer time at the Academic Supercomputer Center (CI TASK) in Gdansk.

\clearpage

\bibliographystyle{ieeetr} 
\bibliography{biblio}

@String{ PR-E   = {Phys. Rev. E}}

@String{ JPC-C  = {J. Phys. Chem. C}}

@String{ PRL  = {Phys. Rev. Lett.}}

@String{ JCP  = {J. Chem. Phys.}}

@article{sivells23,
    author = {Sivells, Tiara and Viswanathan, Pranav and Cyran, Jenée D.},
    title = {Quantification of anion and cation uptake in ice Ih crystals},
    journal = JCP,
    volume = {158},
    number = {13},
    pages = {134507},
    year = {2023},
    month = {04},
    issn = {0021-9606},
    doi = {10.1063/5.0141057},
    url = {https://doi.org/10.1063/5.0141057},
    eprint = {https://pubs.aip.org/aip/jcp/article-pdf/doi/10.1063/5.0141057/16823628/134507\_1\_5.0141057.pdf},
}

@article{wettlaufer99,
  title = {Impurity Effects in the Premelting of Ice},
  author = {Wettlaufer, J.S.},
  journal = PRL,
  volume = {82},
  issue = {12},
  pages = {2516--2519},
  numpages = {4},
  year = {1999},
  month = {Mar},
  publisher = {American Physical Society},
  doi = {10.1103/PhysRevLett.82.2516},
  url = {http://link.aps.org/doi/10.1103/PhysRevLett.82.2516}
}

@article{ribeiro21,
   author = {Ribeiro, Ingrid de Almeida and Koning, Maurice de},
   title = {Grain-Boundary Sliding in Ice Ih: Tribology and Rheology at the
      Nanoscale},
   journal = JPC-C,
   volume = {125},
   number = {1},
   pages = {627-634},
   year = {2021},
   doi = {10.1021/acs.jpcc.0c10032},
   URL = { https://doi.org/10.1021/acs.jpcc.0c10032 },
   eprint = { https://doi.org/10.1021/acs.jpcc.0c10032 }
}

@article{mitsui19,
  title = {Fluctuation spectroscopy of surface melting of ice with and without
impurities},
  author = {Mitsui, Takahisa and Aoki, Kenichiro},
  journal = PR-E,
  volume = {99},
  issue = {1},
  pages = {010801},
  numpages = {5},
  year = {2019},
  month = {Jan},
  publisher = {American Physical Society},
  doi = {10.1103/PhysRevE.99.010801},
  url = {https://link.aps.org/doi/10.1103/PhysRevE.99.010801}
}

@article{perez17,
author="E. P\'erez and Y. Chebude",
title="Analysis of {G}aet'ale, a Hypersaline Pond in {D}anakil Depression ({E}thiopia): New Record for the Most Saline Water Body on Earth.",
journal = "Aquat Geochem.",
volume="23",
pages="109-117",
year="2017",
doi="https://doi.org/10.1007/s10498-017-9312-z"}

@BOOK{petrenko99,
   author = "V. F. Petrenko and R. W. Whitworth",
   title = "Physics of Ice",
   publisher = "Oxford University Press",
   address = {Oxford},
   year = "1999",
   tema = {}
}

@article{zeron21,
author = {Zeron, Ivan M. and Gonzalez, Miguel A. and Errani, Edoardo and Vega, Carlos and Abascal, Jose L. F.},
title = {“In Silico” Seawater},
journal = {Journal of Chemical Theory and Computation},
volume = {17},
number = {3},
pages = {1715-1725},
year = {2021},
doi = {10.1021/acs.jctc.1c00072},
}

@article{blazquez23,
    author = {Blazquez, S. and Conde, M. M. and Vega, C.},
    title = {Scaled charges for ions: An improvement but not the final word for modeling electrolytes in water},
    journal = {The Journal of Chemical Physics},
    volume = {158},
    number = {5},
    pages = {054505},
    year = {2023},
    month = {02},
    issn = {0021-9606},
    doi = {10.1063/5.0136498},
    url = {https://doi.org/10.1063/5.0136498},
}

@article{breton20,
    author = {Le Breton, Guillaume and Joly, Laurent},
    title = {Molecular modeling of aqueous electrolytes at interfaces: Effects of long-range dispersion forces and of ionic charge rescaling},
    journal = {The Journal of Chemical Physics},
    volume = {152},
    number = {24},
    pages = {241102},
    year = {2020},
    month = {06},
    issn = {0021-9606},
    doi = {10.1063/5.0011058},
    url = {https://doi.org/10.1063/5.0011058},
}

@article{blazquez24,
    author = {Blazquez, Samuel and Sedano, Lucia F. and Vega, Carlos},
    title = {On the compatibility of the Madrid-2019 force field for electrolytes with the TIP4P/Ice water model},
    journal = {The Journal of Chemical Physics},
    volume = {161},
    number = {22},
    pages = {224502},
    year = {2024},
    month = {12},
    issn = {0021-9606},
    doi = {10.1063/5.0241233},
    url = {https://doi.org/10.1063/5.0241233},
}

@Article{litman24,
author={Litman, Yair
and Chiang, Kuo-Yang
and Seki, Takakazu
and Nagata, Yuki
and Bonn, Mischa},
title={Surface stratification determines the interfacial water structure of simple electrolyte solutions},
journal={Nature Chemistry},
year={2024},
month={Apr},
day={01},
volume={16},
number={4},
pages={644-650},
issn={1755-4349},
doi={10.1038/s41557-023-01416-6},
url={https://doi.org/10.1038/s41557-023-01416-6}
}

@article{kim08,
author = {Kim, Jun Soo and Yethiraj, Arun},
title = {A Diffusive Anomaly of Water in Aqueous Sodium Chloride Solutions at Low Temperatures},
journal = {The Journal of Physical Chemistry B},
volume = {112},
number = {6},
pages = {1729-1735},
year = {2008},
doi = {10.1021/jp076710+},
URL = {https://doi.org/10.1021/jp076710+},
}

@article{garbacz14,
author = {Garbacz, Piotr and Price, William S.},
title = {1H NMR Diffusion Studies of Water Self-Diffusion in Supercooled Aqueous Sodium Chloride Solutions},
journal = {The Journal of Physical Chemistry A},
volume = {118},
number = {18},
pages = {3307-3312},
year = {2014},
doi = {10.1021/jp501472s},
URL = {https://doi.org/10.1021/jp501472s},
}

@article{bartels12,
  title = {Ice structures, patterns, and processes: A view across the icefields},
  author = {Bartels-Rausch, Thorsten and Bergeron, Vance and Cartwright, Julyan H. E. and Escribano, Rafael and Finney, John L. and Grothe, Hinrich and Guti\'errez, Pedro J. and Haapala, Jari and Kuhs, Werner F. and Pettersson, Jan B. C. and Price, Stephen D. and Sainz-D\'{\i}az, C. Ignacio and Stokes, Debbie J. and Strazzulla, Giovanni and Thomson, Erik S. and Trinks, Hauke and Uras-Aytemiz, Nevin},
  journal = {Rev. Mod. Phys.},
  volume = {84},
  issue = {2},
  pages = {885--944},
  numpages = {0},
  year = {2012},
  month = {May},
  publisher = {American Physical Society},
  doi = {10.1103/RevModPhys.84.885},
  url = {https://link.aps.org/doi/10.1103/RevModPhys.84.885}
}

@Article{magnan21,
author={Magnan, Alexandre K.
and P{\"o}rtner, Hans-Otto
and Duvat, Virginie K. E.
and Garschagen, Matthias
and Guinder, Valeria A.
and Zommers, Zinta
and Hoegh-Guldberg, Ove
and Gattuso, Jean-Pierre},
title={Estimating the global risk of anthropogenic climate change},
journal={Nature Climate Change},
year={2021},
month={Oct},
day={01},
volume={11},
number={10},
pages={879-885},
issn={1758-6798},
doi={10.1038/s41558-021-01156-w},
url={https://doi.org/10.1038/s41558-021-01156-w}
}

@Article{bartels14,
AUTHOR = {Bartels-Rausch, T. and Jacobi, H.-W. and Kahan, T. F. and Thomas, J. L. and Thomson, E. S. and Abbatt, J. P. D. and Ammann, M. and Blackford, J. R. and Bluhm, H. and Boxe, C. and Domine, F. and Frey, M. M. and Gladich, I. and Guzm\'an, M. I. and Heger, D. and Huthwelker, Th. and Kl\'an, P. and Kuhs, W. F. and Kuo, M. H. and Maus, S. and Moussa, S. G. and McNeill, V. F. and Newberg, J. T. and Pettersson, J. B. C. and Roeselov\'a, M. and Sodeau, J. R.},
TITLE = {A review of air–ice chemical and physical interactions (AICI): liquids, quasi-liquids, and solids in snow},
JOURNAL = {Atmospheric Chemistry and Physics},
VOLUME = {14},
YEAR = {2014},
NUMBER = {3},
PAGES = {1587--1633},
URL = {https://acp.copernicus.org/articles/14/1587/2014/},
DOI = {10.5194/acp-14-1587-2014}
}

@article{demmenie25,
  title = {Partial wetting of water on ice},
  author = {Demmenie, Menno and Gorin, Benjamin and Kolpakov, Paul and Smith, Scott and Kellay, Hamid and Bonn, Daniel},
  journal = {Phys. Rev. Fluids},
  volume = {10},
  issue = {5},
  pages = {054002},
  numpages = {9},
  year = {2025},
  month = {May},
  publisher = {American Physical Society},
  doi = {10.1103/PhysRevFluids.10.054002},
  url = {https://link.aps.org/doi/10.1103/PhysRevFluids.10.054002}
}

@article{luengo22,
    author = {Luengo-Márquez, Juan and Izquierdo-Ruiz, Fernando and MacDowell, Luis G.},
    title = {Intermolecular forces at ice and water interfaces: Premelting, surface freezing, and regelation},
    journal = {The Journal of Chemical Physics},
    volume = {157},
    number = {4},
    pages = {044704},
    year = {2022},
    month = {08},
    issn = {0021-9606},
    doi = {10.1063/5.0097378},
    url = {https://doi.org/10.1063/5.0097378},
    eprint = {https://pubs.aip.org/aip/jcp/article-pdf/doi/10.1063/5.0097378/16547392/044704\_1\_online.pdf},
}

@Article{slater19,
author={Slater, Ben
and Michaelides, Angelos},
title={Surface premelting of water ice},
journal={Nature Reviews Chemistry},
year={2019},
month={Mar},
day={01},
volume={3},
number={3},
pages={172-188},
issn={2397-3358},
doi={10.1038/s41570-019-0080-8},
url={https://doi.org/10.1038/s41570-019-0080-8}
}

@article{furukawa87,
title = {Ellipsometric study of the transition layer on the surface of an ice crystal},
journal = {Journal of Crystal Growth},
volume = {82},
number = {4},
pages = {665-677},
year = {1987},
issn = {0022-0248},
doi = {https://doi.org/10.1016/S0022-0248(87)80012-X},
url = {https://www.sciencedirect.com/science/article/pii/S002202488780012X},
author = {Yoshinori Furukawa and Masaki Yamamoto and Toshio Kuroda},
}

@article{bluhm02,
doi = {10.1088/0953-8984/14/8/108},
url = {https://dx.doi.org/10.1088/0953-8984/14/8/108},
year = {2002},
month = {feb},
publisher = {},
volume = {14},
number = {8},
pages = {L227},
author = {Hendrik Bluhm and D Frank Ogletree and Charles S Fadley and Zahid Hussain and Miquel Salmeron},
title = {The premelting of ice studied with photoelectron
spectroscopy},
journal = {Journal of Physics: Condensed Matter},
}

@article{baran24b,
author = {Lukasz Baran and Pablo Llombart and Eva G. Noya and Luis G. MacDowell},
title = {Is it possible to overheat ice? The activated melting of TIP4P/Ice at solid–vapour coexistence.},
journal = {Molecular Physics},
volume = {122},
number = {21-22},
pages = {e2388800},
year = {2024},
publisher = {Taylor \& Francis},
doi = {10.1080/00268976.2024.2388800},
URL = { https://doi.org/10.1080/00268976.2024.2388800 },
eprint = { https://doi.org/10.1080/00268976.2024.2388800 }
}

@article{kling18,
author = {Kling, Tanja and Kling, Felix and Donadio, Davide},
title = {Structure and Dynamics of the Quasi-Liquid Layer at the Surface of Ice from Molecular Simulations},
journal = {The Journal of Physical Chemistry C},
volume = {122},
number = {43},
pages = {24780-24787},
year = {2018},
doi = {10.1021/acs.jpcc.8b07724},
URL = {https://doi.org/10.1021/acs.jpcc.8b07724}
}

@article{louden18,
author = {Louden, Patrick B. and Gezelter, J. Daniel},
title = {Why is Ice Slippery? Simulations of Shear Viscosity of the Quasi-Liquid Layer on Ice},
journal = {The Journal of Physical Chemistry Letters},
volume = {9},
number = {13},
pages = {3686-3691},
year = {2018},
doi = {10.1021/acs.jpclett.8b01339},
    note ={PMID: 29916247},
URL = {https://doi.org/10.1021/acs.jpclett.8b01339},
}

@article{nagata19,
author = {Nagata, Yuki and Hama, Tetsuya and Backus, Ellen H. G. and Mezger, Markus and Bonn, Daniel and Bonn, Mischa and Sazaki, Gen},
title = {The Surface of Ice under Equilibrium and Nonequilibrium Conditions},
journal = {Accounts of Chemical Research},
volume = {52},
number = {4},
pages = {1006-1015},
year = {2019},
doi = {10.1021/acs.accounts.8b00615},
    note ={PMID: 30925035},
URL = {https://doi.org/10.1021/acs.accounts.8b00615}
}

@article{dash06,
  title = {The physics of premelted ice and its geophysical consequences},
  author = {Dash, J. G. and Rempel, A. W. and Wettlaufer, J. S.},
  journal = {Rev. Mod. Phys.},
  volume = {78},
  issue = {3},
  pages = {695--741},
  numpages = {0},
  year = {2006},
  month = {Jul},
  publisher = {American Physical Society},
  doi = {10.1103/RevModPhys.78.695},
  url = {https://link.aps.org/doi/10.1103/RevModPhys.78.695}
}

@article{dash95,
doi = {10.1088/0034-4885/58/1/003},
url = {https://dx.doi.org/10.1088/0034-4885/58/1/003},
year = {1995},
month = {jan},
publisher = {},
volume = {58},
number = {1},
pages = {115},
author = {J G Dash and Haiying Fu and J S Wettlaufer},
title = {The premelting of ice and its environmental consequences},
journal = {Reports on Progress in Physics},
}

@article{wei01,
  title = {Surface Vibrational Spectroscopic Study of Surface Melting of Ice},
  author = {Wei, Xing and Miranda, Paulo B. and Shen, Y. R.},
  journal = {Phys. Rev. Lett.},
  volume = {86},
  issue = {8},
  pages = {1554--1557},
  numpages = {0},
  year = {2001},
  month = {Feb},
  publisher = {American Physical Society},
  doi = {10.1103/PhysRevLett.86.1554},
  url = {https://link.aps.org/doi/10.1103/PhysRevLett.86.1554}
}

@article{faraday60,
author = {Faraday, Michael },
title = {I. Note on regelation},
journal = {Proceedings of the Royal Society of London},
volume = {10},
number = {},
pages = {440-450},
year = {1860},
doi = {10.1098/rspl.1859.0082},
URL = {https://royalsocietypublishing.org/doi/abs/10.1098/rspl.1859.0082},
eprint = {https://royalsocietypublishing.org/doi/pdf/10.1098/rspl.1859.0082}
}

@article{krepelova13,
author = {Křepelová, Ad{\'e}la and Bartels-Rausch, Thorsten and Brown, Matthew A. and Bluhm, Hendrik and Ammann, Markus},
title = {Adsorption of Acetic Acid on Ice Studied by Ambient-Pressure XPS and Partial-Electron-Yield NEXAFS Spectroscopy at 230–240 K},
journal = {The Journal of Physical Chemistry A},
volume = {117},
number = {2},
pages = {401-409},
year = {2013},
doi = {10.1021/jp3102332},
note ={PMID: 23252403},
URL = {://doi.org/10.1021/jp3102332
}
}

@Article{starr11,
author ="Starr, D. E. and Pan, D. and Newberg, J. T. and Ammann, M. and Wang, E. G. and Michaelides, A. and Bluhm, H.",
title  ="Acetone adsorption on ice investigated by X-ray spectroscopy and density functional theory",
journal  ="Phys. Chem. Chem. Phys.",
year  ="2011",
volume  ="13",
issue  ="44",
pages  ="19988-19996",
publisher  ="The Royal Society of Chemistry",
doi  ="10.1039/C1CP21493D",
url  ="http://dx.doi.org/10.1039/C1CP21493D",
}

@article{sanchez17,
author = {M. Alejandra Sánchez  and Tanja Kling  and Tatsuya Ishiyama  and Marc-Jan van Zadel  and Patrick J. Bisson  and Markus Mezger  and Mara N. Jochum  and Jenée D. Cyran  and Wilbert J. Smit  and Huib J. Bakker  and Mary Jane Shultz  and Akihiro Morita  and Davide Donadio  and Yuki Nagata  and Mischa Bonn  and Ellen H. G. Backus },
title = {Experimental and theoretical evidence for bilayer-by-bilayer surface melting of crystalline ice},
journal = {Proceedings of the National Academy of Sciences},
volume = {114},
number = {2},
pages = {227-232},
year = {2017},
doi = {10.1073/pnas.1612893114},
URL = {https://www.pnas.org/doi/abs/10.1073/pnas.1612893114},
eprint = {https://www.pnas.org/doi/pdf/10.1073/pnas.1612893114},
}

@article{qiu18,
author = {Qiu, Yuqing and Molinero, Valeria},
title = {Why Is It So Difficult to Identify the Onset of Ice Premelting?},
journal = {The Journal of Physical Chemistry Letters},
volume = {9},
number = {17},
pages = {5179-5182},
year = {2018},
doi = {10.1021/acs.jpclett.8b02244},
    note ={PMID: 30149705},

URL = {https://doi.org/10.1021/acs.jpclett.8b02244}
}

@article{murata16,
author = {Kenichiro Murata  and Harutoshi Asakawa  and Ken Nagashima  and Yoshinori Furukawa  and Gen Sazaki },
title = {Thermodynamic origin of surface melting on ice crystals},
journal = {Proceedings of the National Academy of Sciences},
volume = {113},
number = {44},
pages = {E6741-E6748},
year = {2016},
doi = {10.1073/pnas.1608888113},
URL = {https://www.pnas.org/doi/abs/10.1073/pnas.1608888113},
eprint = {https://www.pnas.org/doi/pdf/10.1073/pnas.1608888113},
}

@article{jungwirth06,
author = {Jungwirth, Pavel and Tobias, Douglas J.},
title = {Specific Ion Effects at the Air/Water Interface},
journal = {Chemical Reviews},
volume = {106},
number = {4},
pages = {1259-1281},
year = {2006},
doi = {10.1021/cr0403741},
    note ={PMID: 16608180},

URL = {https://doi.org/10.1021/cr0403741}
}

@article{asakawa16,
author = {Harutoshi Asakawa  and Gen Sazaki  and Ken Nagashima  and Shunichi Nakatsubo  and Yoshinori Furukawa },
title = {Two types of quasi-liquid layers on ice crystals are formed kinetically},
journal = {Proceedings of the National Academy of Sciences},
volume = {113},
number = {7},
pages = {1749-1753},
year = {2016},
doi = {10.1073/pnas.1521607113},
URL = {https://www.pnas.org/doi/abs/10.1073/pnas.1521607113},
eprint = {https://www.pnas.org/doi/pdf/10.1073/pnas.1521607113},
}

@Article{mcneill12,
AUTHOR = {McNeill, V. F. and Grannas, A. M. and Abbatt, J. P. D. and Ammann, M. and Ariya, P. and Bartels-Rausch, T. and Domine, F. and Donaldson, D. J. and Guzman, M. I. and Heger, D. and Kahan, T. F. and Kl\'an, P. and Masclin, S. and Toubin, C. and Voisin, D.},
TITLE = {Organics in environmental ices: sources, chemistry, and impacts},
JOURNAL = {Atmospheric Chemistry and Physics},
VOLUME = {12},
YEAR = {2012},
NUMBER = {20},
PAGES = {9653--9678},
URL = {https://acp.copernicus.org/articles/12/9653/2012/},
DOI = {10.5194/acp-12-9653-2012}
}

@article{mcneill06,
author = {V. Faye McNeill  and Thomas Loerting  and Franz M. Geiger  and Bernhardt L. Trout  and Mario J. Molina },
title = {Hydrogen chloride-induced surface disordering on ice},
journal = {Proceedings of the National Academy of Sciences},
volume = {103},
number = {25},
pages = {9422-9427},
year = {2006},
doi = {10.1073/pnas.0603494103},
URL = {https://www.pnas.org/doi/abs/10.1073/pnas.0603494103},
eprint = {https://www.pnas.org/doi/pdf/10.1073/pnas.0603494103},
}

@article{elbaum93,
title = {Optical study of surface melting on ice},
journal = {Journal of Crystal Growth},
volume = {129},
number = {3},
pages = {491-505},
year = {1993},
issn = {0022-0248},
doi = {https://doi.org/10.1016/0022-0248(93)90483-D},
url = {https://www.sciencedirect.com/science/article/pii/002202489390483D},
author = {Michael Elbaum and S.G. Lipson and J.G. Dash},
}

@article{conde08,
    author = {Conde, M. M. and Vega, C. and Patrykiejew, A.},
    title = {The thickness of a liquid layer on the free surface of ice as obtained from computer simulation},
    journal = {The Journal of Chemical Physics},
    volume = {129},
    number = {1},
    pages = {014702},
    year = {2008},
    month = {07},
    issn = {0021-9606},
    doi = {10.1063/1.2940195},
    url = {https://doi.org/10.1063/1.2940195},
    eprint = {https://pubs.aip.org/aip/jcp/article-pdf/doi/10.1063/1.2940195/15415428/014702\_1\_online.pdf},
}

@article{llombart20,
  title = {Rounded Layering Transitions on the Surface of Ice},
  author = {Llombart, Pablo and Noya, Eva G. and Sibley, David N. and Archer, Andrew J. and MacDowell, Luis G.},
  journal = {Phys. Rev. Lett.},
  volume = {124},
  issue = {6},
  pages = {065702},
  numpages = {6},
  year = {2020},
  month = {Feb},
  publisher = {American Physical Society},
  doi = {10.1103/PhysRevLett.124.065702},
  url = {https://link.aps.org/doi/10.1103/PhysRevLett.124.065702}
}

@Article{berrens22,
author ="Berrens, Margaret L. and Bononi, Fernanda C. and Donadio, Davide",
title  ="Effect of sodium chloride adsorption on the surface premelting of ice",
journal  ="Phys. Chem. Chem. Phys.",
year  ="2022",
volume  ="24",
issue  ="35",
pages  ="20932-20940",
publisher  ="The Royal Society of Chemistry",
doi  ="10.1039/D2CP02277J",
url  ="http://dx.doi.org/10.1039/D2CP02277J",
}

@article{kahan14,
author = {Kahan, Tara F. and Wren, Sumi N. and Donaldson, D. James},
title = {A Pinch of Salt Is All It Takes: Chemistry at the Frozen Water Surface},
journal = {Accounts of Chemical Research},
volume = {47},
number = {5},
pages = {1587-1594},
year = {2014},
doi = {10.1021/ar5000715},
    note ={PMID: 24785086},

URL = {https://doi.org/10.1021/ar5000715}
}

@Article{singh01,
author={Singh, H.
and Chen, Y.
and Staudt, A.
and Jacob, D.
and Blake, D.
and Heikes, B.
and Snow, J.},
title={Evidence from the Pacific troposphere for large global sources of oxygenated organic compounds},
journal={Nature},
year={2001},
month={Apr},
day={01},
volume={410},
number={6832},
pages={1078-1081},
issn={1476-4687},
doi={10.1038/35074067},
url={https://doi.org/10.1038/35074067}
}

@article{hundait17,
author = {Hudait, Arpa and Allen, Michael T. and Molinero, Valeria},
title = {Sink or Swim: Ions and Organics at the Ice–Air Interface},
journal = {Journal of the American Chemical Society},
volume = {139},
number = {29},
pages = {10095-10103},
year = {2017},
doi = {10.1021/jacs.7b05233},
    note ={PMID: 28658949},

URL = {https://doi.org/10.1021/jacs.7b05233}
}

@Article{peterson18,
AUTHOR = {Peterson, A. K.},
TITLE = {Observations of brine plumes below melting Arctic sea ice},
JOURNAL = {Ocean Science},
VOLUME = {14},
YEAR = {2018},
NUMBER = {1},
PAGES = {127--138},
URL = {https://os.copernicus.org/articles/14/127/2018/},
DOI = {10.5194/os-14-127-2018}
}

@article{crabeck19,
author = {Crabeck, O. and Galley, R. J. and Mercury, L. and Delille, B. and Tison, J.-L. and Rysgaard, S.},
title = {Evidence of Freezing Pressure in Sea Ice Discrete Brine Inclusions and Its Impact on Aqueous-Gaseous Equilibrium},
journal = {Journal of Geophysical Research: Oceans},
volume = {124},
number = {3},
pages = {1660-1678},
doi = {https://doi.org/10.1029/2018JC014597},
url = {https://agupubs.onlinelibrary.wiley.com/doi/abs/10.1029/2018JC014597},
eprint = {https://agupubs.onlinelibrary.wiley.com/doi/pdf/10.1029/2018JC014597},
year = {2019}
}

@article{ingolf75,
title={The Formation of Brine Drainage Features in Young Sea Ice},
volume={14},
DOI={10.3189/S0022143000013460},
number={70},
journal={Journal of Glaciology},
author={Ingolf Eide, Lars and Martin, Seelye},
year={1975}, pages={137–154}
}

@article{junge01, 
title={A microscopic approach to investigate bacteria under in situ conditions in sea-ice samples}, 
volume={33}, 
DOI={10.3189/172756401781818275}, 
journal={Annals of Glaciology}, 
author={Junge, Karen and Krembs, Christopher and Deming, Jody and Stierle, Aaron and Eicken, Hajo}, 
year={2001}, pages={304–310}
}

@article{shcherbina03,
author = {Andrey Y. Shcherbina  and Lynne D. Talley  and Daniel L. Rudnick },
title = {Direct Observations of North Pacific Ventilation: Brine Rejection in the Okhotsk Sea},
journal = {Science},
volume = {302},
number = {5652},
pages = {1952-1955},
year = {2003},
doi = {10.1126/science.1088692},
URL = {https://www.science.org/doi/abs/10.1126/science.1088692},
eprint = {https://www.science.org/doi/pdf/10.1126/science.1088692},
}

@article{abascal05,
  author =        {J. L. F. Abascal and E. Sanz and
                   R. Garc{\'i}a Fernandez and C. Vega},
  journal =       {J. Chem. Phys.},
  pages =         {234511},
  title =         {A potential model for the study of ices and amorphous
                   water: {TIP4P/Ice}},
  volume =        {122},
  year =          {2005}
}

@article{gough11,
title = {Laboratory studies of perchlorate phase transitions: Support for metastable aqueous perchlorate solutions on Mars},
journal = {Earth and Planetary Science Letters},
volume = {312},
number = {3},
pages = {371-377},
year = {2011},
issn = {0012-821X},
doi = {https://doi.org/10.1016/j.epsl.2011.10.026},
url = {https://www.sciencedirect.com/science/article/pii/S0012821X11006212},
author = {R.V. Gough and V.F. Chevrier and K.J. Baustian and M.E. Wise and M.A. Tolbert},
}

@article{toner14,
title = {The formation of supercooled brines, viscous liquids, and low-temperature perchlorate glasses in aqueous solutions relevant to Mars},
journal = {Icarus},
volume = {233},
pages = {36-47},
year = {2014},
issn = {0019-1035},
doi = {https://doi.org/10.1016/j.icarus.2014.01.018},
url = {https://www.sciencedirect.com/science/article/pii/S0019103514000499},
author = {J.D. Toner and D.C. Catling and B. Light},
}

@Article{aristov97,
author={Aristov, Yu. I.
and Di Marco, G.
and Tokarev, M. M.
and Parmon, V. N.},
title={Selective water sorbents for multiple applications, 3. CaCl2 solution confined in micro- and mesoporous silica gels: Pore size effect on the ``solidification-melting'' diagram},
journal={Reaction Kinetics and Catalysis Letters},
year={1997},
month={May},
day={01},
volume={61},
number={1},
pages={147-154},
issn={1588-2837},
doi={10.1007/BF02477527},
url={https://doi.org/10.1007/BF02477527}
}

@article{christenson01,
doi = {10.1088/0953-8984/13/11/201},
url = {https://dx.doi.org/10.1088/0953-8984/13/11/201},
year = {2001},
month = {mar},
publisher = {},
volume = {13},
number = {11},
pages = {R95},
author = {Hugo K Christenson},
title = {Confinement effects on freezing and melting},
journal = {Journal of Physics: Condensed Matter},
}

@article{cho02,
author = {Cho, H. and Shepson, P. B. and Barrie, L. A. and Cowin, J. P. and Zaveri, R.},
title = {NMR Investigation of the Quasi-Brine Layer in Ice/Brine Mixtures},
journal = {The Journal of Physical Chemistry B},
volume = {106},
number = {43},
pages = {11226-11232},
year = {2002},
doi = {10.1021/jp020449+},

URL = {https://doi.org/10.1021/jp020449+}
}

@article{richardson76,
title={Phase Relationships in Sea Ice as a Function of Temperature},
volume={17},
DOI={10.3189/S0022143000013770},
number={77},
journal={Journal of Glaciology},
author={Richardson, C.},
year={1976},
pages={507–519}
}

@article{buffo20,
author = {Buffo, J. J. and Schmidt, B. E. and Huber, C. and Walker, C. C.},
title = {Entrainment and Dynamics of Ocean-Derived Impurities Within Europa's Ice Shell},
journal = {Journal of Geophysical Research: Planets},
volume = {125},
number = {10},
pages = {e2020JE006394},
doi = {https://doi.org/10.1029/2020JE006394},
url = {https://agupubs.onlinelibrary.wiley.com/doi/abs/10.1029/2020JE006394},
eprint = {https://agupubs.onlinelibrary.wiley.com/doi/pdf/10.1029/2020JE006394},
note = {e2020JE006394 2020JE006394},
year = {2020}
}

@article{vance21,
author = {Vance, Steven D. and Journaux, Baptiste and Hesse, Marc and Steinbrügge, Gregor},
title = {The Salty Secrets of Icy Ocean Worlds},
journal = {Journal of Geophysical Research: Planets},
volume = {126},
number = {1},
pages = {e2020JE006736},
doi = {https://doi.org/10.1029/2020JE006736},
url = {https://agupubs.onlinelibrary.wiley.com/doi/abs/10.1029/2020JE006736},
eprint = {https://agupubs.onlinelibrary.wiley.com/doi/pdf/10.1029/2020JE006736},
note = {e2020JE006736 2020JE006736},
year = {2021}
}

@article{mohler05,
author = {Möhler, O. and Büttner, S. and Linke, C. and Schnaiter, M. and Saathoff, H. and Stetzer, O. and Wagner, R. and Krämer, M. and Mangold, A. and Ebert, V. and Schurath, U.},
title = {Effect of sulfuric acid coating on heterogeneous ice nucleation by soot aerosol particles},
journal = {Journal of Geophysical Research: Atmospheres},
volume = {110},
number = {D11},
pages = {},
doi = {https://doi.org/10.1029/2004JD005169},
url = {https://agupubs.onlinelibrary.wiley.com/doi/abs/10.1029/2004JD005169},
eprint = {https://agupubs.onlinelibrary.wiley.com/doi/pdf/10.1029/2004JD005169},
year = {2005}
}

@article{popp04,
author = {Popp, P. J. and Gao, R. S. and Marcy, T. P. and Fahey, D. W. and Hudson, P. K. and Thompson, T. L. and Kärcher, B. and Ridley, B. A. and Weinheimer, A. J. and Knapp, D. J. and Montzka, D. D. and Baumgardner, D. and Garrett, T. J. and Weinstock, E. M. and Smith, J. B. and Sayres, D. S. and Pittman, J. V. and Dhaniyala, S. and Bui, T. P. and Mahoney, M. J.},
title = {Nitric acid uptake on subtropical cirrus cloud particles},
journal = {Journal of Geophysical Research: Atmospheres},
volume = {109},
number = {D6},
pages = {},
doi = {https://doi.org/10.1029/2003JD004255},
url = {https://agupubs.onlinelibrary.wiley.com/doi/abs/10.1029/2003JD004255},
eprint = {https://agupubs.onlinelibrary.wiley.com/doi/pdf/10.1029/2003JD004255},
year = {2004}
}

@article{heckendorn09,
doi = {10.1088/1748-9326/4/4/045108},
url = {https://dx.doi.org/10.1088/1748-9326/4/4/045108},
year = {2009},
month = {nov},
publisher = {},
volume = {4},
number = {4},
pages = {045108},
author = {Heckendorn, P and Weisenstein, D and Fueglistaler, S and Luo, B P and Rozanov, E and Schraner, M and Thomason, L W and Peter, T},
title = {The impact of geoengineering aerosols on stratospheric temperature and ozone},
journal = {Environmental Research Letters},
}

@article{solomon11,
author = {S. Solomon  and J. S. Daniel  and R. R. Neely  and J.-P. Vernier  and E. G. Dutton  and L. W. Thomason },
title = {The Persistently Variable “Background” Stratospheric Aerosol Layer and Global Climate Change},
journal = {Science},
volume = {333},
number = {6044},
pages = {866-870},
year = {2011},
doi = {10.1126/science.1206027},
URL = {https://www.science.org/doi/abs/10.1126/science.1206027},
eprint = {https://www.science.org/doi/pdf/10.1126/science.1206027},
}

@article{lammps22,
  author =        {A. P. Thompson and H. M. Aktulga and R. Berger and
                   D. S. Bolintineanu and W. M. Brown and P. S. Crozier and
                   P. J. in 't Veld and A. Kohlmeyer and S. G. Moore and
                   T. D. Nguyen and R. Shan and M. J. Stevens and
                   J. Tranchida and C. Trott and S. J. Plimpton},
  journal =       {Comp. Phys. Comm.},
  pages =         {108171},
  title =         {{LAMMPS} - a flexible simulation tool for
                   particle-based materials modeling at the atomic,
                   meso, and continuum scales},
  volume =        {271},
  year =          {2022},
  doi =           {10.1016/j.cpc.2021.108171}
}

@article{kamberaj05,
    author = {Kamberaj, H. and Low, R. J. and Neal, M. P.},
    title = {Time reversible and symplectic integrators for molecular dynamics simulations of rigid molecules},
    journal = {The Journal of Chemical Physics},
    volume = {122},
    number = {22},
    pages = {224114},
    year = {2005},
    month = {06},
    issn = {0021-9606},
    doi = {10.1063/1.1906216},
    url = {https://doi.org/10.1063/1.1906216},
}

@article{darden93,
  author =        {T. A. Darden and D. York and L. Pedersen},
  journal =       {J. Chem. Phys.},
  pages =         {10089--10092},
  title =         {Particle Mesh Ewald: an {N log(N) Method for Ewald
                   Sums in Large Systems}},
  volume =        {98},
  year =          {1993}
}

@article{conde17,
  author =        {Conde, M. M. and Rovere, M. and Gallo, P.},
  journal =       {The Journal of Chemical Physics},
  month =         {12},
  number =        {24},
  pages =         {244506},
  title =         {{High precision determination of the melting points
                   of water TIP4P/2005 and water TIP4P/Ice models by the
                   direct coexistence technique}},
  volume =        {147},
  year =          {2017},
  doi =           {10.1063/1.5008478},
  issn =          {0021-9606},
  url =           {https://doi.org/10.1063/1.5008478}
}

@article{nguyen15,
  author =        {Andrew H. Nguyen and Valeria Molinero},
  journal =       {J. Phys. Chem. B},
  number =        {29},
  pages =         {9369 - 9376},
  title =         {Identification of Clathrate Hydrates, Hexagonal Ice,
                   Cubic Ice, and Liquid Water in Simulations: the
                   CHILL+ Algorithm},
  volume =        {119},
  year =          {2015}
}

@article{daivis94,
    author = {Daivis, Peter J. and Evans, Denis J.},
    title = {Comparison of constant pressure and constant volume nonequilibrium simulations of sheared model decane},
    journal = {The Journal of Chemical Physics},
    volume = {100},
    number = {1},
    pages = {541-547},
    year = {1994},
    month = {01},
    issn = {0021-9606},
    doi = {10.1063/1.466970},
    url = {https://doi.org/10.1063/1.466970},
    eprint = {https://pubs.aip.org/aip/jcp/article-pdf/100/1/541/19306281/541\_1\_online.pdf},
}

@article{conde17b,
    author = {Conde, M. M. and Rovere, M. and Gallo, P.},
    title = {Spontaneous NaCl-doped ice at seawater conditions: focus on the mechanisms of ion inclusion},
    journal = {Physical Chemistry Chemical Physics},
    volume = {19},
    pages = {9566-9574},
    year = {2017},
    month = {03},
    doi = {10.1039/C7CP00665A},
}

@article{sedano24,
author = {L. F. Sedano and C. Vega and E. G. Noya},
title = {Isothermal compressibility and water self-diffusion coefficient in supercooled salt aqueous solutions using the Madrid-2019 model},
journal = {Molecular Physics},
volume = {122},
number = {21-22},
pages = {e2418311},
year = {2024},
publisher = {Taylor \& Francis},
doi = {10.1080/00268976.2024.2418311},
URL = { https://doi.org/10.1080/00268976.2024.2418311},
}

@article{baran22,
author = {Łukasz Baran  and Pablo Llombart  and Wojciech Rżysko  and Luis G. MacDowell },
title = {Ice friction at the nanoscale},
journal = {Proceedings of the National Academy of Sciences},
volume = {119},
number = {49},
pages = {e2209545119},
year = {2022},
doi = {10.1073/pnas.2209545119},
URL = {https://www.pnas.org/doi/abs/10.1073/pnas.2209545119},
eprint = {https://www.pnas.org/doi/pdf/10.1073/pnas.2209545119},
}

@misc{macdowell25,
      title={The Key Physics of Ice Premelting}, 
      author={Luis G. MacDowell},
      year={2025},
      eprint={2509.13221},
      archivePrefix={arXiv},
      primaryClass={cond-mat.soft},
      url={https://arxiv.org/abs/2509.13221}, 
}

@article{baran25,
    author = {Baran, \L ukasz and Dicu-Gohoreanu, Cosmin A. and MacDowell, Luis G.},
    title = {How important is the dielectric constant in water modeling? Evaluation of the performance of the TIP4P/$\varepsilon$ force field and its compatibility with the Joung–Cheatham NaCl model},
    journal = {The Journal of Chemical Physics},
    volume = {163},
    number = {5},
    pages = {054504},
    year = {2025},
    month = {08},
    issn = {0021-9606},
    doi = {10.1063/5.0283754},
    url = {https://doi.org/10.1063/5.0283754},
}

@Article{dickson13,
author={Dickson, James L.
and Head, James W.
and Levy, Joseph S.
and Marchant, David R.},
title={Don {J}uan Pond, {A}ntarctica: Near-surface {C}a{Cl}2-brine feeding Earth's most saline lake and implications for Mars},
journal={Scientific Reports},
year={2013},
month={Jan},
day={30},
volume={3},
number={1},
pages={1166},
issn={2045-2322},
doi={10.1038/srep01166},
url={https://doi.org/10.1038/srep01166}
}

@article{lamas22,
    author = {Lamas, Cintia P. and Vega, Carlos and Noya, Eva G.},
    title = {Freezing point depression of salt aqueous solutions using the Madrid-2019 model},
    journal = {The Journal of Chemical Physics},
    volume = {156},
    number = {13},
    pages = {134503},
    year = {2022},
    month = {04},
    issn = {0021-9606},
    doi = {10.1063/5.0085051},
    url = {https://doi.org/10.1063/5.0085051},
}

@article{oakes90,
author = {Charles S Oakes and Robert J Bodnar and John M Simonson},
title = {The system NaCl-CaCl2-H2O: I. The ice liquidus at 1 atm total pressure},
journal = {Geochimica et Cosmochimica Acta},
volume = {54},
number = {3},
pages = {603-610},
year = {1990},
issn = {0016-7037},
doi = {https://doi.org/10.1016/0016-7037(90)90356-P},
url = {https://www.sciencedirect.com/science/article/pii/001670379090356P},
}

@article{nada05,
title = {Anisotropy in growth kinetics at interfaces between proton-disordered hexagonal ice and water: A molecular dynamics study using the six-site model of H2O},
journal = {Journal of Crystal Growth},
volume = {283},
number = {1},
pages = {242-256},
year = {2005},
issn = {0022-0248},
doi = {https://doi.org/10.1016/j.jcrysgro.2005.05.057},
url = {https://www.sciencedirect.com/science/article/pii/S0022024805006482},
author = {H. Nada and Y. Furukawa},
}

\includepdf[pages=-]{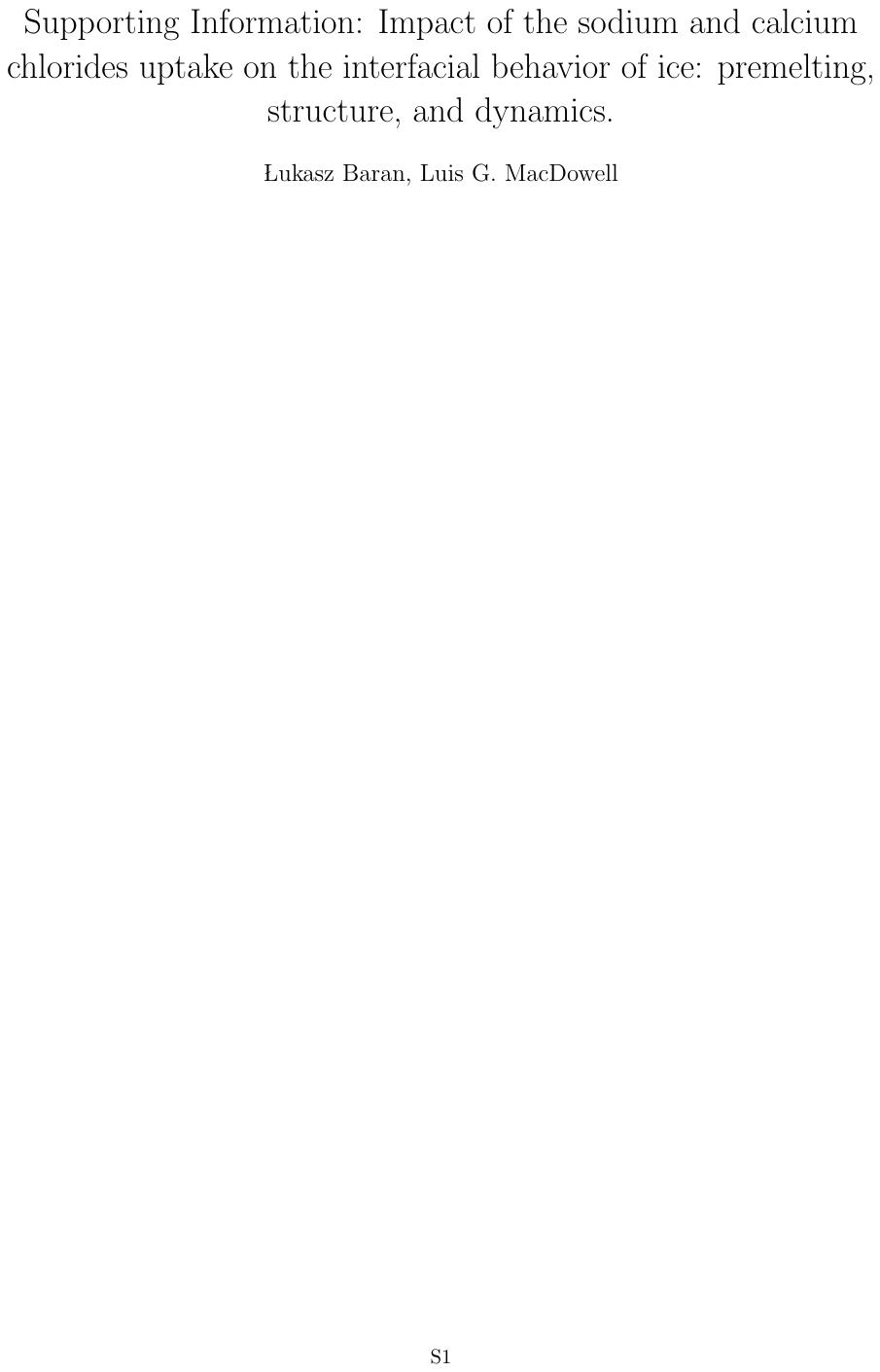}

\end{document}